\documentclass[prb,aps,floatfix,superscriptaddress,twocolumn]{revtex4-1}
\usepackage{amsmath,amsfonts,amssymb,graphics,graphicx,color,bm}
\usepackage{mathtools}
\usepackage{braket}
\usepackage{bbold}
\usepackage[normalem]{ulem}
\usepackage{overpic}
\usepackage{tabularx}
\usepackage{booktabs}
\usepackage{algorithm}
\usepackage[noend]{algpseudocode}
\usepackage{hyperref}
\usepackage[english]{babel}

\newcommand{\vecbf}[1]{{\bf #1}}

\begin{document}

\title{Quantum state tomography on a plaquette in the 2D Hubbard model}
\date{\today}
\author{Stephan Humeniuk}
\email{smhumeniuk@iphy.ac.cn}
\affiliation{Beijing National Laboratory for Condensed Matter Physics and Institute of Physics, Chinese Academy of Sciences, Beijing 100190, China}

\begin{abstract}

Motivated by recent quantum gas microscope experiments for fermions in optical lattices, 
we present proof of principle calculations showing that it is possible to obtain the complete information 
about the quantum state on a small subsystem from equilibrium determinantal quantum Monte Carlo simulations. 
Both diagonal (in the occupation number basis) and off-diagonal elements of the reduced density matrix are calculated 
for a square plaquette, which is embedded in a much larger system of the two-dimensional Hubbard model, both 
at half filling and in the doped case. 
The diagonalization of the reduced density matrix is done by exploiting 
the point group symmetry and particle number conservation,
which allows to attach symmetry labels to its eigenvalues. 

Knowledge of the probabilities of plaquette occupation number configurations is useful 
for meticulous benchmarking of quantum gas microscope experiments.
As the quantum state on the plaquette is exact and self-consistently embedded in an exact, correlated bath,
the present approach connects to various cluster approximation techniques.

\end{abstract}
\maketitle


\section{Introduction}

Quantum state tomography refers to the task of reconstructing the full quantum state 
of a system from measurements, which, by definition, is a task that scales exponentially in the 
system size.
Tomography has been used to characterize small quantum systems consisting of a few qubits,
such as trapped ion chains \cite{Haeffner2005},
molecules in nuclear magnetic resonance experiments \cite{TaoXin2017},
superconducting circuits \cite{Baur2012}, and photonic systems \cite{Schwemmer2014,JunGao2018}. 
More efficient quantum tomography techniques based on compressed sensing \cite{Gross2010}
and matrix product states \cite{Cramer2010} have been proposed,
facilitating the reconstruction based on incomplete data \cite{Schwemmer2014,Riofrio2017,Steffens2015}.

For cold atoms in optical lattices emulating strongly correlated solid state models
such a global characterization of the state is neither feasible 
nor meaningful. Still, the local state on a subsystem of sites, encoded by its reduced 
density matrix $\rho_A = \text{Tr}_{B=\bar{A}}\left(\rho \right)$,
which is given by the partial trace of the global density matrix $\rho = e^{-\beta \hat{H}} / Z$
over the complement system $B$, can provide valuable information if subjected to 
a tomographic measurement. 
Here, we present proof of principle determinantal quantum Monte Carlo (DQMC) 
calculations for a square plaquette of the Fermi-Hubbard model which is embedded in a much larger system. 

Cold atoms experiments with quantum gas microscopes have reached a regime where 
local antiferromangetic correlations in the two-dimensional 
Fermi-Hubbard model and the effect of doping away from half filling 
can be explored \cite{Mazurenko2017, Cheuk2016, Mitra2017, Koepsell2018}.
In particular, signatures of polarons \cite{Koepsell2018, Chiu2018} have been reported. 

The finite-temperature phase diagram of the 2D Fermi Hubbard model 
for intermediate interaction strength \cite{Siimkovic2017}
is very challenging due to the fermionic sign problem, which is aggravated exponentially by
decreasing temperature. The temperatures currently realized in
fermionic quantum gas microscopes experiments are still within reach 
of numerically exact DQMC simulations,
and detailed comparison of full particle 
number distribution functions \cite{Humeniuk2017} has been made.

There are numerous ramifications that motivate computing the equilibrium probabilities 
of individual microstates, i.e. the diagonal elements of $\rho_A$.
By superimposing a specific experimental measurement protocol, 
the effect of parity projection \cite{Parsons2016}, 
i.e. the inability to dinstinguish doubly occupied sites and holes in the imaging process,
on the experimentally observed particle configurations can be assessed.

Remarkably, a scheme for measuring the second moment of the density matrix,
its purity $\text{Tr}\left( \rho_A^2 \right)$, without exponential effort
has been demonstrated in bosonic cold atoms experiments \cite{Daley2012,Islam2015}
and generalized to fermionic systems \cite{Pichler2013}. 
Measurement protocols for accessing the full entanglement 
spectrum in cold atoms systems have also been proposed \cite{Pichler2016}.
In view of this and  recent progress in using machine learning with neural networks 
to reconstruct a full quantum state of bosonic systems~\footnote{The sign structure 
of fermionic wave functions makes the application of 
machine learning approaches to fermionic systems more involved.}
from a limited number of experimental measurements \cite{Torlai2018,Torlai2019}, 
numerical access to all (diagonal and off-diagonal) elements of $\rho_A$ is of potential interest. 
Alternative numerical approaches relying 
on the replica representation of R\'{e}nyi entropies \cite{Chung2014} $\text{Tr}\left[ \rho_A^n\right]$ 
require complicated modifications in the topology of the simulation cell, whereas 
the  brute force numerical scheme described here
does not affect the core of the DQMC algorithm and all diagonal and off-diagonal elements
of the reduced density matrix, as well as its eigenvalues, the \emph{entanglement spectrum}, can be obtained. 
However, computational and memory 
resources that grow exponentially with the subsystem size limit the latter 
to maximally $N_s=9$ sites~\footnote{The computational
complexity of DQMC for simulating the total system of $N$ sites is $\sim \beta N^3$ and 
can be completely decoupled from the costly ``exact diagonalization'' inside each HS sample 
if the single-particle Green's functions are saved on disk for every HS configuration
(or after a number of Monte Carlo steps proportional to the autocorrelation time).
This requires several hundred GB of hard disk memory per parameter set $(\beta, U)$.}.

From a methodological point of view, there are connections with various types 
of numerical cluster approaches. 
Eqs.~\eqref{eq:QST_projector_diag} and \eqref{eq:QST_projector_offdiag}  below
give the exact state on a cluster that is self-consistently embedded in a \emph{correlated} bath
and can be used to compare with computationally less expensive methods
that solve the cluster system exactly, but treat the bath degrees of freedom only approximately. 
In Refs.~\cite{Udagawa2010, Udagawa2015}, using an auxiliary-field quantum Monte Carlo 
solver in the context of cellular dynamical mean-field theory, the 
entanglement spectrum was computed for a triangular plaquette of the kagome Hubbard model,
revealing an emergent composite degree of freedom due to geometric frustration. 
The method used there is very similar in spirit to the one presented in this work. 

Finally, a strong motivation for studying the detailed structure of a local quantum state comes 
from the phenomenology of the high-temperature phase of the repulsive Hubbard model 
(or more generally of high-$T_c$ superconductors) where for temperatures $\beta \le 4 -5  $, where DQMC
simulations are still possible due to a mild sign problem \cite{Iglovikov2015}, a pseudogap 
develops in the single-particle spectral function \cite{Preuss1995}. 
The pseudogap in the \emph{attractive} Hubbard model is well understood in terms of local  bound pairs 
of fermions without long-range phase coherence, the gap being associated 
with the binding energy of the pair. A natural question is whether similar preformed objects 
are responsible for the pseudogap observed in the \emph{repulsive} Hubbard model \cite{Preuss1995}
(or more generally in the normal state of high-$T_c$ superconductors)
and which signatures of correlated phases are contained in the local density matrix \cite{Huber2019}.

The structure of this paper is as follows. 
In Sect.~\ref{sec:Borns_rule}, we describe the algorithm for projecting the reduced 
density matrix $\rho_A$ on a subsystem $A$ from the global density matrix which is 
sampled in the DQMC procedure. 
The symmetries of the reduced density matrix on a square plaquette and the transformation to the 
irreducible representation basis is discussed in Sect.~\ref{sec:symmetries_rhoA}.
In Sect.~\ref{sec:error_bars} the issue of error bars is addressed. Finally,
Sect.~\ref{sec:results} contains the results for the tomographic reconstruction of $\rho_A$
on a square plaquette of the Hubbard model: At half filling, we present the diagonal 
and off-diagonal elements of $\rho_A$ as a function of Hubbard interaction $U/t$
for both high and low temperatures. Away from half-filling, where the computational cost of 
DQMC simulations is affected by the sign problem, we show the doping dependence of $\rho_A$ for  
$U/t=6$ and a relatively high temperature of $T = 0.25 t$, 
which corresponds to the lowest temperature achieved 
so far in fermionic cold atoms experiments \cite{Mazurenko2017}. Sect.~\ref{sec:conclusion}
concludes with an outlook.

\section{Born's rule for many-body states}
\label{sec:Borns_rule}

The Hamiltonian studied in the following is that of the single-band Hubbard model 
\begin{align}
 \mathcal{H} = - &t \sum_{\langle i, j \rangle, \sigma=\uparrow, \downarrow} \left( c_{i,\sigma}^{\dagger} c_{j,\sigma} + \text{H.c.} \right) + \mu \sum_{i=1}^{N}(n_{i,\uparrow} + n_{i,\downarrow}) \nonumber\\
 + &U \sum_{i=1}^{N} n_{i,\uparrow} n_{i,\downarrow},
 \label{eq:Hubbard_Hamiltonian}
\end{align}
where $c_{i,\sigma}^{\dagger}$  creates a fermion with spin $\sigma$ at site $i$,
and \mbox{$n_{i,\sigma} = c_{i,\sigma}^{\dagger} c_{i,\sigma}$}.
Here, $t$ is the hopping matrix element between nearest neighbours $\langle i, j \rangle$, 
$\mu$ is the chemical potential, and $U$ is the on-site repulsive ($U>0$) or attractive ($U<0$) interaction.
The partition sum 
\begin{equation}
 Z = \text{Tr}\left( e^{-\beta \mathcal{H}} \right)
\end{equation}
is sampled with the determinantal quantum Monte Carlo method \cite{Blankenbecler1981, Loh1992, Assaad2002}. 
We briefly sketch the essential conceptual steps in the derivation of this procedure,
referring to the exhaustive literature (see e.g.~Refs.~\cite{Loh1992, Assaad2002}) for more details. 
After discretizing inverse temperature $\beta = N_{\tau} \Delta \tau$ into $N_{\tau}$ imaginary time slices 
and separating the single-body kinetic term in the Hamiltonian from the two-body interaction term 
via a Trotter-Suzuki decomposition, a Hubbard-Stratonovich (HS) transformation is applied 
to the interaction term converting it into a single-particle term which is coupled 
to a fluctuating space- and imaginary time-dependent potential given by auxiliary field variables. Thanks to 
the HS transformation 
the partition sum contains only exponentials of bilinear (i.e. free) fermionic operators. Then, 
the free fermions can be integrated out for each auxiliary field configuration 
using the well-known formula for the 
grand-canonical fermionic trace, which results in 
\begin{align}
 Z &= \sum_{\{ \vecbf{s} \}} \prod_{\sigma = {\uparrow, \downarrow}} \text{det} \left( \mathbb{1} + B^{\sigma}_{N_{\tau}} B^{\sigma}_{N_{\tau}-1} \cdots B^{\sigma}_1\right) \\
   &= \sum_{\{ \vecbf{s} \}} w^{\uparrow}_{\{ \vecbf{s}\}} w^{\downarrow}_{\{ \vecbf{s}\}}.
\end{align}
Here, $B_{l}^{\sigma} \equiv e^{-\Delta \tau V_l^{\sigma}(\{\vecbf{s}_l\})} e^{- \Delta \tau K}$ is 
an $N \times N$ matrix representation of the single-particle propagator at time slice $l$ after HS transformation,
with $V_l^{\sigma}(\{ \vecbf{s}_l \})$ denoting the potential term and $K$ the kinetic term 
of the single-particle Hamiltonian after HS transformation \cite{Assaad2002}, while 
$\{ \vecbf{s} \} \equiv \{ s_{i,l}\}_{i=1, \ldots,N; l=1,\ldots, N_{\tau}}$
is the space-time configuration of auxiliary-field variables, which is sampled with 
a Monte Carlo technique. 

It can be shown (see e.g. Ref.~\cite{Assaad2002}) that the  
weight $w^{\sigma}_{\{ \vecbf{s} \}}$ is given by the inverse determinant of the 
equal-time single-particle Green's function
\begin{align}
 &\left[G^{\sigma}_{\{ \vecbf{s}\}} (\tau = l \Delta \tau)\right]_{i,j} \equiv \langle c_{i,\sigma} c_{j,\sigma}^{\dagger} \rangle_{\{ \vecbf{s} \}}  \nonumber \\
 &= \left[\left( \mathbb{1} + B^{\sigma}_{l} B^{\sigma}_{l-1} \cdots B^{\sigma}_{1} B^{\sigma}_{N_{\tau}} \cdots B^{\sigma}_{l+1} \right)^{-1} \right]_{i,j},
\end{align}
which constitutes the central quantity of the DQMC algorithm. Furthermore, for the Hubbard model
the HS transformation can be chosen such that the weight of an auxiliary-field configuration 
$\{ \vecbf{s} \}$ factorizes between spin species \cite{Assaad2002}. 

Thus, loosely speaking, the DQMC framework
consists in computing a large sum over free fermions systems
in varying external potentials: It suffices to compute any quantity (in any single-particle basis)
for free fermions and average over Monte Carlo samples. This is a huge conceptual simplification
compared to path integral methods. The generality of the free fermion decomposition \cite{Grover2013}
allows to carry the measurement part of the DQMC algorithm to its extreme
by calculating the full quantum state on a small subsystem $A$, that is 
all elements of the reduced density matrix $\rho_A$.
This amounts to performing exact diagonalization inside the measurement part of the 
Monte Carlo procedure.

The diagonal and off-diagonal elements of the reduced density matrix in the Fock basis can be written as 
\begin{align}
\langle \alpha | \rho | \alpha \rangle &= \sum_{\{\vecbf{s}\}}\text{Tr}\left( \rho_{\{\vecbf{s}\}}\, \hat{\Pi}_{\alpha} \right)  \label{eq:QST_projector_diag} \\
\langle \beta | \rho | \alpha \rangle  &= \sum_{\{\vecbf{s}\}}\text{Tr}\left( \rho_{\{\vecbf{s}\}}\, \hat{\Xi}_{\alpha \rightarrow \beta} \right), \quad \alpha \ne \beta. \label{eq:QST_projector_offdiag}
\end{align}
Here, $\hat{\Pi}_{\alpha} = |\alpha \rangle \langle \alpha |$ are projectors onto individual Fock states
\begin{equation}
|\alpha\rangle = |\alpha_{\uparrow} \rangle \otimes  |\alpha_{\downarrow} \rangle = |n_{1, \uparrow},n_{1, \downarrow};n_{2, \uparrow},n_{2, \downarrow}; \ldots; n_{N_s, \uparrow},n_{N_s, \downarrow} \rangle 
\label{eq:tensor_Fock_state}
\end{equation}
on the substem $A$ with $N_s$ sites~\footnote{Translational invariance can be used 
to accumulate additional statistics by displacing the subsystem
repeatedly. However, for the calculations presented in the following
for a single plaquette, the location of the plaquette was fixed.} 
and $\hat{\Xi}_{\alpha \rightarrow \beta} = |\beta \rangle \langle \alpha |$ is a transition operator
between two Fock states $|\alpha\rangle$
and $|\beta \rangle$. $\rho_{\{\bf s\}}$ is the global density matrix of the 
free fermion system with auxiliary field configuration $\{ \vecbf{s} \}$, which is sampled via Monte Carlo. 

In a state of a non-interacting Fermi system, Wick's theorem applied to a product of $n$ pairings 
of fermionic operators results in the determinant formula 
\begin{equation}
 \left\langle \left( c_{i_1} c_{j_1}^{\dagger} \right) \left( c_{i_2} c_{j_2}^{\dagger} \right) \cdots \left( c_{i_n} c_{j_n}^{\dagger}\right)  \right\rangle_0
  = \det \left( G_{i_{\alpha} j_{\beta}}^{(0)} \right)
  \label{eq:Wicks_theorem_general}
\end{equation}
with $ \alpha, \beta=1,\ldots, n$, 
where the equal-time Green's function of the non-interacting Fermi system
is \mbox{$G_{i_{\alpha} j_{\beta}}^{(0)}(\tau=0) = \langle c_{i_{\alpha}}(\tau=0) c_{j_{\beta}}^{\dagger}(0) \rangle_0$}. 
This formula is the basis for evaluating the expectation values of the projectors and transition 
operators in Eqs.~\eqref{eq:QST_projector_diag} and \eqref{eq:QST_projector_offdiag}
and thus extracting the elements of the reduced density matrix from an equilibrium state,
which in the DQMC framework is encoded in a sum over free fermion systems parametrized by auxiliary 
field configurations $\{ \vecbf{s} \}$.

Since for one auxiliary field configuration $\{\vecbf{s}\}$ the reduced density matrix factorizes between spin species,
\begin{equation}
\begin{split}
 \left( \langle \beta_{\uparrow} | \otimes \langle \beta_{\downarrow} | \right) \rho_{A, \{\vecbf{s}\}}^{\uparrow} 
  \otimes \rho_{A, \{\vecbf{s}\}}^{\downarrow} \left( | \alpha_{\uparrow} \rangle \otimes | \alpha_{\downarrow} \rangle \right) \\
 =  \langle \alpha_{\uparrow} | \rho_{A, \{ \vecbf{s} \}}^{\uparrow} | \beta_{\uparrow} \rangle 
    \langle \alpha_{\downarrow} | \rho_{A, \{ \vecbf{s} \}}^{\downarrow} | \beta_{\downarrow} \rangle,
\end{split}    
 \end{equation}
at most $2\times 2^{N_{s}} \times 2^{N_{s}}$ matrix elements need to be computed 
to express all $4^{N_{s}} \times 4^{N_{s}}$ elements of $\rho_{A,\{\vecbf{s}\}}$.
(In this crude estimate we have disregarded the block diagonal structure of $\rho_{A,\{\vecbf{s}\}}^{\sigma}$
with respect to particle number $N^{\sigma}_A$ which reduces the size of the largest particle number block for one 
spin species 
to \mbox{$\begin{pmatrix} N_s \\ \lfloor N_s/2\rfloor \end{pmatrix} \cdot \begin{pmatrix} N_s \\ \lfloor N_s/2 \rfloor \end{pmatrix}$}
with $\lfloor x \rfloor$ denoting the largest integer that is smaller than $x$.)
Therefore, the limiting factor is the memory requirement for storing all elements of $\rho_{A,\{\vecbf{s}\}}$
for Monte Carlo averaging, rather than the computation of individual elements. 
The presence of point group symmetry operations which leave subsystem $A$
invariant or spin inversion symmetry leads to an additional block diagonal 
structure of $\rho_A$ (see Sect.~\ref{sec:particle_number_conservation}).

In the following, we discuss the  computation for a single spin component, 
thereby dropping all spin indices in the notation.  
We use hats to distinguish the number operator $\hat{n}_i = c_i^{\dagger} c_i$ from the occupation number $n_i$
and write 
\begin{align}
 \hat{\Pi}_{\alpha} = \prod_{i \in N_s} \left[ \hat{n}_i n_i + (1 - \hat{n}_i)(1-n_i) \right] \nonumber \\
                    = \prod_{i \text{ occupied}} \hat{n_i} \prod_{j \text{ unoccupied}} (1 - \hat{n}_j).
\label{eq:projector_Fock}                    
\end{align}
The transition operator 
$\hat{\Xi}_{\alpha \rightarrow \beta}$ can be written as 
\begin{equation}
 \hat{\Xi}_{\alpha \rightarrow \beta} = \hat{T}_{\alpha \rightarrow \beta}\, \hat{\Pi}_{\alpha},
\end{equation}
where $\hat{\Pi}_{\alpha}$ projects onto the Fock state $| \alpha\rangle$, which is then converted into $|\beta \rangle$
by a combination of appropriately chosen creation and annihilation operators
\begin{equation}
 \hat{T}_{\alpha \rightarrow \beta} = (-1)^p \left(\prod_{\substack{c=N^{+}\\i_1 < i_2 < \cdots < i_{N^{+}}}}^{1} c_{i_c}^{\dagger}\right) \left( \prod_{\substack{a=N^{-}\\j_1 < j_2 < \cdots < j_{N^{-}}}}^{1} c_{j_a} \right).
 \label{eq:T_alpha2beta}
\end{equation}
The sequences of site indices $\mathcal{I}^{+} = \{i_1,i_2, \ldots, i_{N^{+}}\}$ 
and $\mathcal{I}^{-} = \{j_1,j_2, \ldots, j_{N^{-}}\}$, ordered according to the 
chosen fermion ordering, denote the lattice sites where the $N^{+}$ creation and $N^{-}$ annihilation operators 
must act to convert $|\alpha \rangle$ into $|\beta \rangle$.
Since $\rho_A$ is block diagonal with respect to the total particle number 
$N_{A,\sigma} = \sum_{i \in A} n_{i,\sigma}$
for each spin species $\sigma$ (see Sect.~\ref{sec:particle_number_conservation}), 
there must be as many creation as annihilation operators and $N^{+} = N^{-}$.
The fermionic phase
\begin{equation}
 (-1)^{p} = \prod_{i_c=1}^{N^{+}} e^{i \pi \sum_{i_c < l < N_s} n_{l}^{(\gamma)}} 
            \prod_{j_a=1}^{N^{-}} e^{i \pi \sum_{j_a < k < N_s} n_{k}^{(\alpha)}}
\label{eq:minus1_p}                        
\end{equation}
with $\vecbf{n}^{(\alpha)}$ the vector of occupation numbers for state $|\alpha\rangle$
and $\vecbf{n}^{(\gamma)}$ for state $|\gamma \rangle \equiv \prod_{a=N^{-}}^{1} \, c_{j_a} | \alpha \rangle$
ensures that $\hat{T}_{\alpha \rightarrow \beta} |\alpha \rangle = | \beta \rangle$.
Given that $N^{+}=N^{-}$, one can bring the operator product in Eq.~\eqref{eq:T_alpha2beta} into 
the paired form as it appears on the left-hand side
of Eq.~\eqref{eq:Wicks_theorem_general} by commuting fermionic operators. 
From this an additional phase factor arises:
\begin{equation}
 (-1)^{p^{\prime}} = (-1)^{\sum_{i=1}^{N^{+}} i} = (-1)^{\frac{N^{+}}{2}(N^{+}+1)}.
 \label{eq:minus1_p_prime}
\end{equation}

With regard to a practical implementation for the evaluation 
of the expectation values in the right-hand side of Eqs.~\eqref{eq:QST_projector_diag} and \eqref{eq:QST_projector_offdiag} 
for one particular auxiliary field configuration $\{\vecbf{s}\}$, a few remarks are in order:
The fact that an occupied site (``occ.'' in the code listing in Appendix \ref{app:pseudocode}) 
is represented by a projector of the form $\hat{n}_{i} = (1 - c_{i} c^{\dagger}_i)$
(for unoccupied sites $1 - \hat{n}_{i} = c_{i} c^{\dagger}_i$) means that the expectation value of the total 
projector onto a Fock state Eq.~\eqref{eq:projector_Fock}, when multiplied out,  is a sum
of terms which can be written as a binary tree for the occupied sites where each leaf
is of the form of Eq.~\eqref{eq:Wicks_theorem_general}. The branches of the binary tree need to 
be summed over to obtain the projector  Eq.~\eqref{eq:projector_Fock}.
In combination with $\hat{T}_{\alpha \rightarrow \beta}$, the projector $\hat{\Pi}_{\alpha}$ 
needs only be realized on sites that are unaffected (``ua" in the code listing in Appendix \ref{app:pseudocode}) 
by the hopping operators in $\hat{T}_{\alpha \rightarrow \beta}$
since the hopping operators already guarantee that occupation number states $\alpha^{\prime} \ne \alpha$
are eliminated by the action of $\hat{T}_{\alpha \rightarrow \beta}$.

\begin{figure}
 \includegraphics[width=1.0\linewidth]{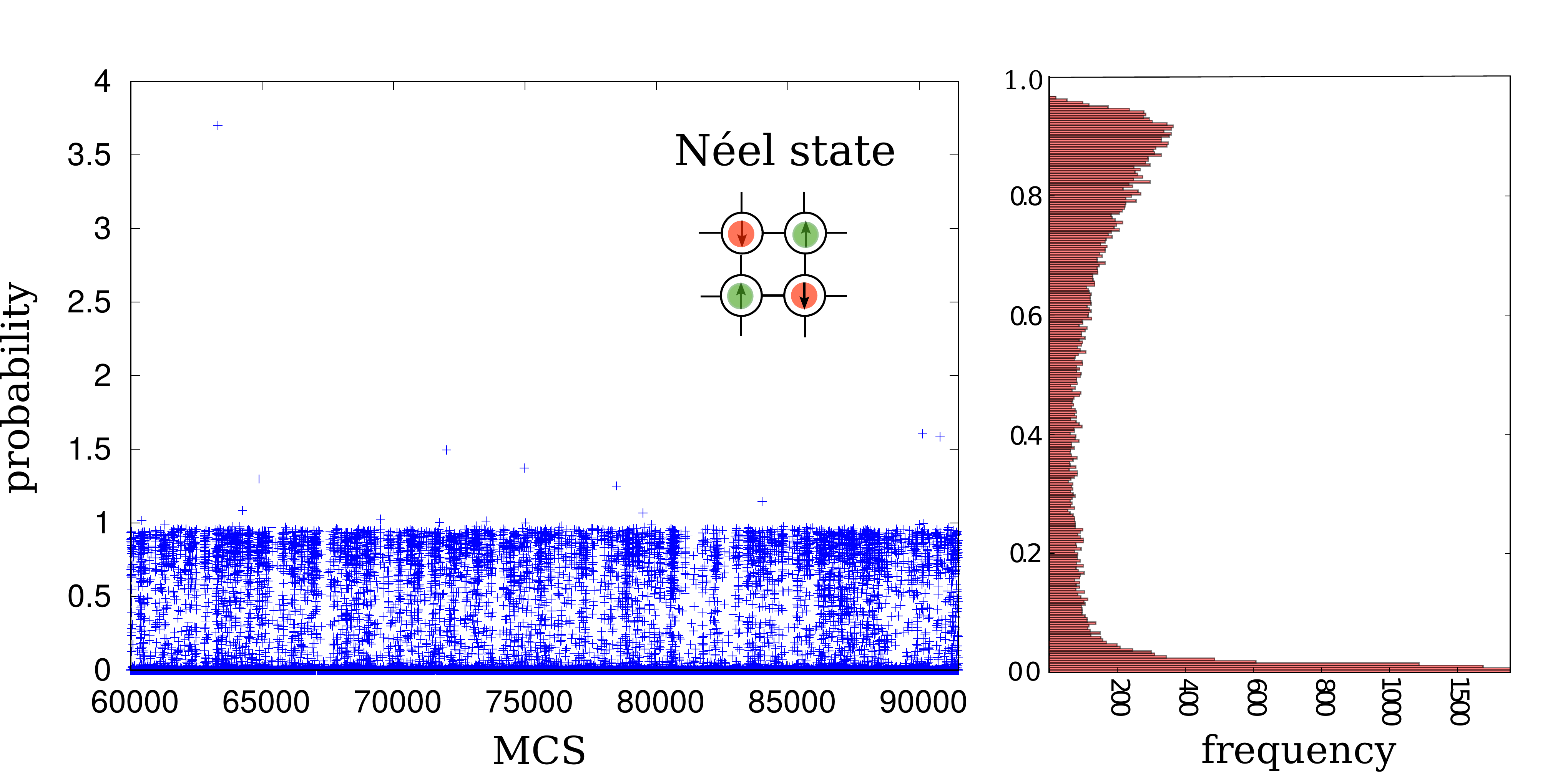}
 \caption[Monte Carlo timeseries of the probability for one of the two plaquette N\'{e}el states.]
 {Monte Carlo timeseries of the probability $P$ for 
 one of the two plaquette N\'{e}el states. Note the violation of $P \le 1$ for 
 rare outliers. $U/t=7.2, \beta t = 4$, system size $L\times L$ with $L=12$.}
 \label{fig:freqNeel}
\end{figure}

The algorithm for computing $\langle \beta | \rho_A | \alpha \rangle_{\{ \vecbf{s} \}}$ between the occupation number states $|\alpha \rangle$
and $|\beta \rangle$ for a single Hubbard-Stratonovich configuration $\{ \vecbf{s} \}$ is summarized
in the pseudocode listing in Appendix ~\ref{app:pseudocode}, where the main task consists in collecting the 
correct row and column indices for the submatrices that enter the determinant formula Eq.~\eqref{eq:Wicks_theorem_general}.
The final result for a matrix element of $\rho_A$ in the interacting system 
is obtained by summing over all Hubbard-Stratonovich configurations $\{ \vecbf{s} \}$.
As an illustration, the Monte Carlo timeseries of the diagonal element 
of $\rho_{A}$ corresponding to one of the two N\'{e}el states on a plaquette 
is displayed in Fig.~\ref{fig:freqNeel}. 
There are rare outliers that exceed to maximum probability of $1$. 
The histogram of probabilities on the right 
is slightly bimodal but smooth; the mean value is around $P_{\text{N\'{e}el}} \approx 0.12$.
For a projective measurement
one would expect a binary distribution of probabilities 
with only the probabilities $0$ or $1$ appearing.

\section{Symmetries of the reduced density matrix}
\label{sec:symmetries_rhoA}

The form of the global density matrix 
\begin{equation}
 \rho = \frac{1}{Z} e^{-\beta(H - \mu N)} \quad \text{with} \quad Z=\text{Tr}\left( e^{-\beta H} \right) \nonumber 
\end{equation}
implies that a symmetry operation represented by a unitary operator $U$ which
obeys the commutation relation $[H - \mu N,U] = 0$, is trivially also a symmetry
of the global density matrix:
\begin{equation}
U^{\dagger} \rho U = \rho. 
\label{eq:rho_invariant}
\end{equation}
In the following it is discussed how symmetries of the Hamiltonian affect the block diagonal structure 
of the reduced density matrix $\rho_A$ of a subsystem $A$. 
Details on the exact diagonalization of the Hubbard model by means of symmetries 
are discussed in Refs.~\cite{Fano1992,Noce1996}; an analytical diagonalization exploiting all symmetries 
was carried out in Ref.~\cite{Schumann2002} for a single plaquette of the Hubbard model 
and in Ref.~\cite{Kuns2011} for a plaquette of the $t-J$ model.

\subsection{Particle number conservation}
\label{sec:particle_number_conservation}

The eigenstates of the reduced density matrix retain their good quantum
numbers when the corresponding operator of the total system is a direct
sum of the operators of its subsystems \cite{Chen1996}. 
This is the case for the particle number $\hat{N}_{\sigma} = \hat{N}_{A,\sigma} + \hat{N}_{B,\sigma}$
(or the $z$ component of the total spin $\hat{M} = \hat{N}_{\uparrow} - \hat{N}_{\downarrow}$)
and consequently $\rho_A$ is block diagonal with respect 
to the quantum numbers $N_{A,\sigma} = \sum_{\vecbf{i} \in A} n_{\vecbf{i},\sigma}$.
For brevity, we denote particle number sectors on the subsystem $A$ 
in the following by $(N_{\uparrow}, N_{\downarrow})$,
where $N_{\sigma} \equiv N_{A,\sigma}$.

\subsection{Lattice symmetry: point group $D_4$}
We focus in the following on the point group symmetry of the square lattice, the 
non-Abelian dihedral group $D_4$ with $h=8$ group elements 
\begin{equation}
D_4 = \{\mathcal{E},  C_{2z}, C_{4z}, C_{4z}^{-1}, C_{2x}, C_{2y}, C_{2xy}, C_{2x\bar{y}} \}
\end{equation}
comprising the identity $\mathcal{E}$ and (assuming that the square is lying in the $x-y$ plane)
rotations by $\pi$ around the $x$, $y$ and $z$ axes, $ C_{2x}, C_{2y}$ and $C_{2z}$,
rotations by $\pi$ around the diagonal lines $x = y$ and $x = -y$, $C_{2xy}$ 
and $C_{2x\bar{y}}$,
and clockwise and counterclockwise rotations by $\pi/2$ 
around the $z$ axis, $C_{4z}$ and $C_{4z}^{-1}$.

The group $D_4$ has five irreducible representations, four one-dimensional representations
with the Mulliken symbols $A_1$, $A_2$, $B_1$, and $B_2$, and one two-dimensional representation $E$. 
For later reference, they are listed in Tab.~\ref{tab:irrep_D4} together with their symmetries.
(see \cite{TinkhamGroupTheory} for the character table and the irreducible representation matrices of $E$).
\begin{table}
 \centering
 \begin{ruledtabular} 
 \begin{tabular}{@{}lccc@{}}
 \toprule 
 \toprule
 Mulliken symbol & dim. & basis function & symmetry   \\
 \midrule
 $A_1$  &  1   & $(x^2 + y^2)\cdot z^2$ &  $s$                     \\
 $A_2$  &  1   & $(x^2 + y^2)\cdot z$ &  $s$                     \\
 $B_1$  &  1   & $x^2-y^2$ &  $d_{x^2 - y^2}$               \\
 $B_2$  &  1   & $xy$  &  $d_{xy}$        \\
 $E  $  &  2   & $x,y$ &  $p_x, p_y$              \\
 \bottomrule  
 \end{tabular}
 \end{ruledtabular}
\caption{Irreducible representations of the symmetry group $D_4$.}
\label{tab:irrep_D4}
\end{table}

We denote by $\mathcal{L}_{A(B)}$ the geometric object consisting of the lattice sites in subsystem $A$ (or in its complement $B)$.
Consider the subgroup $\tilde{\mathcal{G}}$ of lattice symmetry operations that can be written as
\begin{equation}
 R = R^{(A)} R^{(B)} \quad \text{with} \quad R^{(A)} \mathcal{L}_A = \mathcal{L}_A \, \text{and}\, R^{(B)} \mathcal{L}_B = \mathcal{L}_B,
 \label{eq:joint_symmetry_AB}
\end{equation}
where $R^{(A)}$ ($R^{(B)}$) acts only on sites in $A$ ($B$).
By
\begin{equation}
 P_R = P_R^{(A)} \otimes P_R^{(B)}
\end{equation}
we denote the corresponding operator 
that acts onto wave functions in second quantization rather than
lattice sites (cf. Eq.~\eqref{eq:Wigner_convention} below).
Then the general invariance of the global $\rho$ under all elements $R$ of the 
point group $\mathcal{G}$
\begin{equation}
 P_R^{\dagger} \, \rho \, P_R = \rho
\end{equation}
implies for the subgroup $\tilde{\mathcal{G}}$ of elements $R^{\prime}$
which can be written in the specific form Eq.~\eqref{eq:joint_symmetry_AB} that 
\begin{align}
 \text{Tr}_B \left( {P_{R^{\prime}}^{(A)}}^{\dagger} \otimes {P_{R^{\prime}}^{(B)}}^{\dagger} \, \rho \, P_{R^{\prime}}^{(A)} \otimes P_{R^{\prime}}^{(B)} \right) &= \text{Tr}_B \left(\rho\right) \equiv \rho_A \nonumber \\
 \Rightarrow \, {P_{R^{\prime}}^{(A)}}^{\dagger} \text{Tr}_B \left( {P_{R^{\prime}}^{(B)}}^{\dagger} \, \rho \, {P_{R^{\prime}}^{(B)}} \right) P_{R^{\prime}}^{(A)} &= \rho_A \nonumber \\
 \Rightarrow \, {P_{R^{\prime}}^{(A)}}^{\dagger} \rho_A P_{R^{\prime}}^{(A)} &= \rho_A.
\end{align}
In the last step the basis independence of the trace operation and the definition of the reduced density matrix was used. 
Thus, $\rho_A$ is invariant under all joint lattice symmetries $R^{\prime}$ of $\mathcal{L}_A$ and $\mathcal{L}_B$ that map each
subset of lattice sites separately back onto itself according to Eq.~\eqref{eq:joint_symmetry_AB}. 
If either $\mathcal{L}_A$ or $\mathcal{L}_B$ has reduced symmetry (e.g. a square plaquette embedded in a rectangular 
system or a rectangular plaquette inside a square system), then only 
the largest common symmetry subgroup $\tilde{\mathcal{G}}$ of  the point group $\mathcal{G}$ 
is inherited by $\rho_A$. 
An illustrative example is shown in the inset of Fig.~\ref{fig:plaquette_corner}
where the plaquette $\mathcal{L}_A$ possesses the full symmetry of the square, but due to its location
at the corner of a system with open boundary conditions the complement lattice $\mathcal{L}_B$
is only invariant under $C_{2 x\bar{y}}$, which is reflected in the symmetries of the diagonal elements of $\rho_A$ 
(see main panel of Fig.~\ref{fig:plaquette_corner}).

Particle number sectors with $N_{\uparrow} = N_{\downarrow}$ can be further decomposed 
according whether the states are even or odd under spin the inversion symmetry $S$ (see 
The symmetry group of these sectors is $D_{4h} = D_4 \times S$, allowing 
a finer symmetry labelling (see Appendix \ref{app:spin_inversion_symm}).

Once all basis vectors $| \phi_{i \lambda}^{(n)}\rangle$ have been constructed,
the reduced density matrix can be transformed from the occupation to the representation basis via 
\begin{equation}
 \rho_A^{(\text{representation})} = S^{\dagger} \rho_A^{(\text{occupation})} S
 \label{eq:occ_to_rep}
\end{equation}
with the transformation matrix $S_{\alpha, (n,i,\lambda)} = \langle \alpha | \phi_{i \lambda}^{(n)}\rangle$.
Here, $| \phi_{i \lambda}^{(n)}\rangle$ is the basis state corresponding to the $\lambda$-th 
copy of $i$-th row of the $n$-th irreducible representation. The group theoretical techniques 
for constructing these basis vectors are detailed in Appendix \ref{app:projection_operator_method}.

\begin{figure}
 \includegraphics[width=0.49\linewidth]{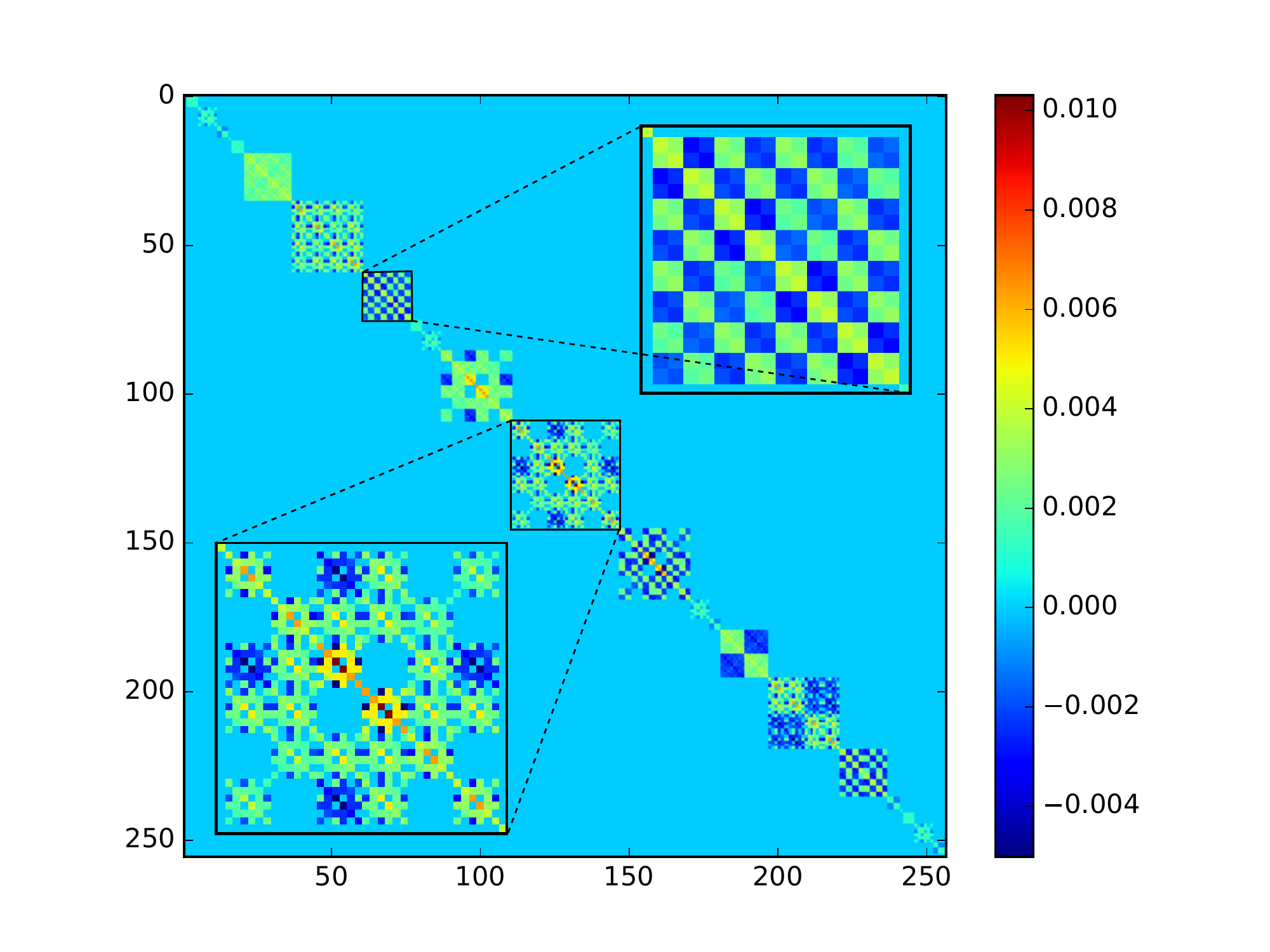} \includegraphics[width=0.49\linewidth]{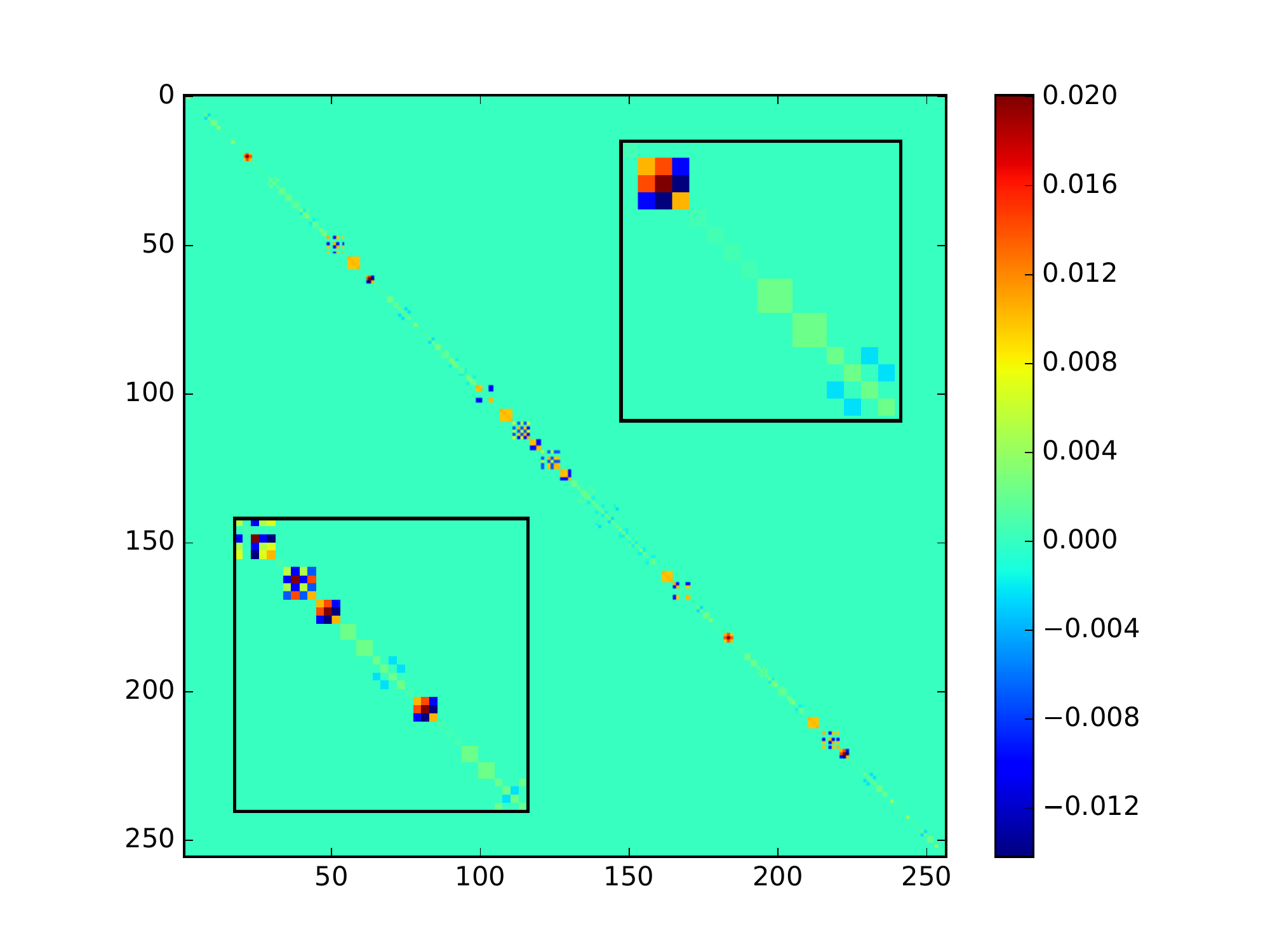}
 \includegraphics[width=0.49\linewidth]{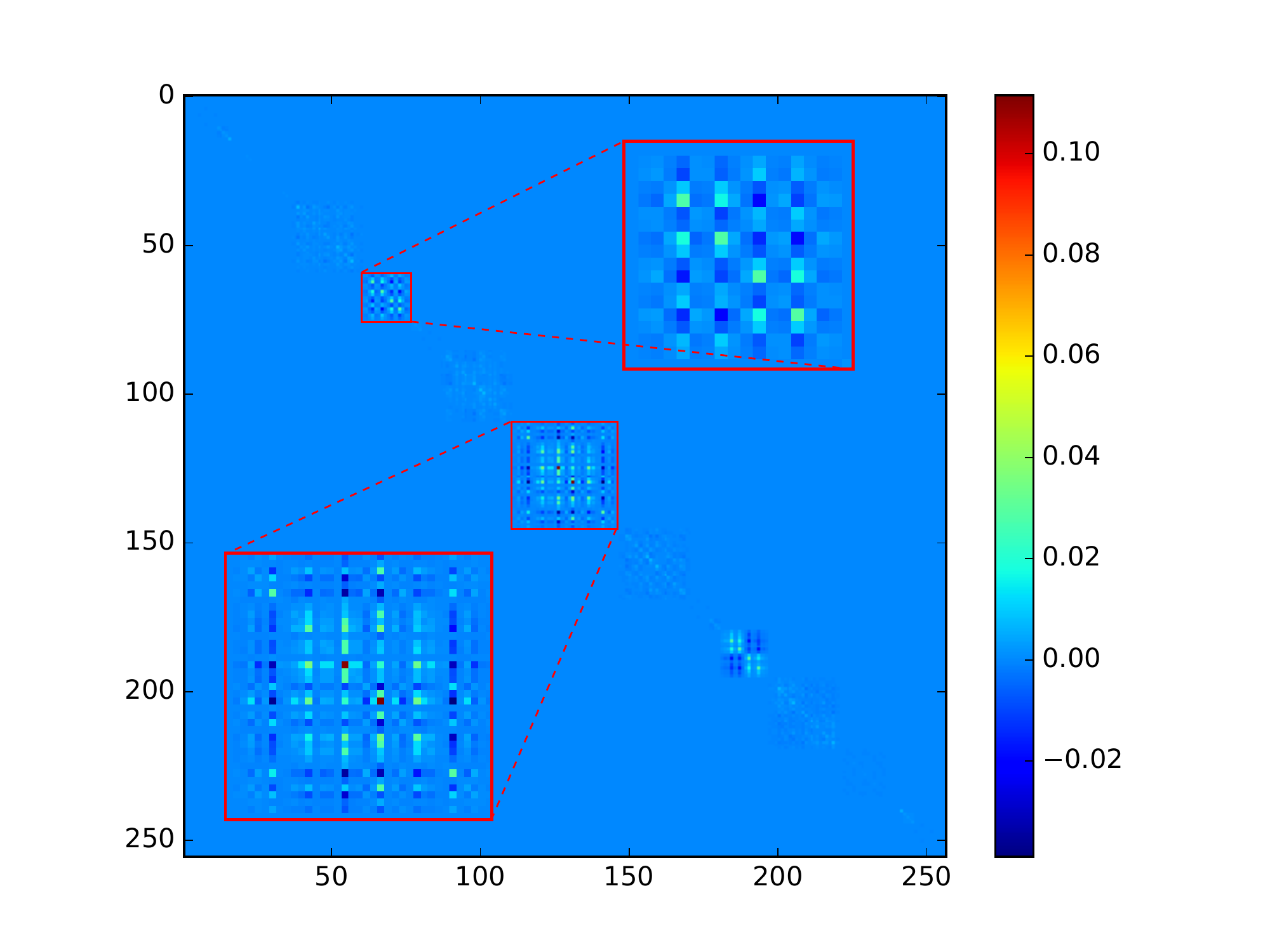} \includegraphics[width=0.49\linewidth]{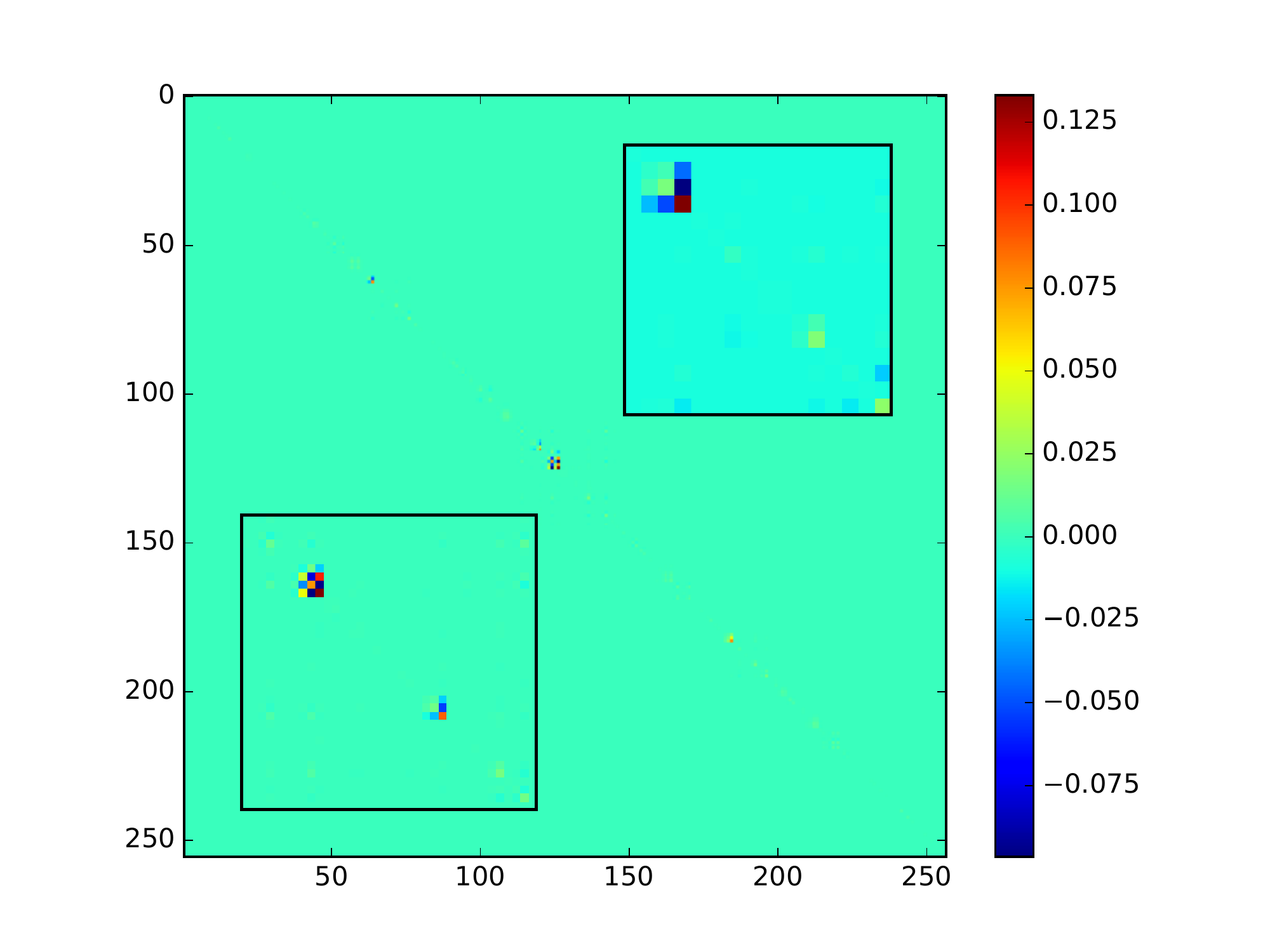}
 \caption[Transformation of the reduced density matrix $\rho_{A=\Box}$ from the occupation basis to the irreducible representation basis.]
 {Transformation of the reduced density matrix $\rho_{A=\Box}$ from the occupation basis (left column) to the representation basis (right column) of the symmetry group $D_{4h}$. First row $U=0, \beta t=4$;
 second row $U=6t,\mu=3t, \beta t=4$. The total system size is $L \times L$ with $L=12$. The insets 
 show the enlarged particle number blocks $(N_{\uparrow}=2, N_{\downarrow}=2)$ and $(N_{\uparrow}=1, N_{\downarrow}=3)$ in the lower left and upper right corner, respectively. 
 Note that the block $(2,2)$ in the inset is decomposed with respect to the irreducible representations of $D_{4h}$ 
 while in the main panel only the symmetry group $D_4$ is used.}
 \label{fig:rhoA_occ_to_rep}
\end{figure}

As can be seen from Fig.~\ref{fig:rhoA_occ_to_rep}, 
when written in the irreducible representation basis, 
the reduced density matrix $\rho_A$ does not have non-vanishing matrix elements between
states of different symmetry and acquires a block diagonal form. 
More importantly, we are in a position to attach symmetry labels to the eigenvalues of $\rho_A$.

\section{Error bars}
\label{sec:error_bars}
Due to statistical fluctuations $\rho_A$ cannot be perfectly Hermitian,
however, the deviations from Hermiticity, $\Delta_H = \rho_A - \rho_A^{\dagger}$, are smaller than the error bars of the 
corresponding off-diagonal elements, and $\rho_A$ is found to be normalized, $\text{Tr}(\rho_A) =1 \pm \varepsilon$,
with an inaccuracy ranging from $\varepsilon \lesssim  10^{-5}$ ($0 \le U \le 4$) 
to $\varepsilon \approx (1 - 3) \times  10^{-2}$ (large $U$, $4 \le U \le 10$) for $\beta t = 4$. 
For low temperatures the inaccuracy is slightly larger ($\epsilon \approx 3 \times 10^{-2}$)
for all values of $U$ (see insets in Fig.~\ref{fig:eigvals_rhoA_allU} below).
Furthermore, $\rho_A$ is positive semi-definite within statistical uncertainty, as required for a valid density matrix.
Error bars have been obtained with the bootstrap method, in which the matrix diagonalization
is repeated $\sim 10^3$ times, each time adding Gaussian noise with a standard
deviation of the size of the deviation from Hermiticity $|\langle \beta | \Delta_H | \alpha \rangle|$
to each matrix element $\langle \beta | \rho_A |\alpha \rangle$. 
The well-resolved symmetry-related degeneracies seen in Figs.~\ref{fig:eigvals_rhoA_allU} (a) and (b) below 
indicate that this error analysis is sound~\footnote{Due to the normalization all elements of $\rho_A$ have correlations 
among them so that adding Gaussian noise idependently to all elements cannot be entirely correct.}. 

\section{QST for a plaquette in the Hubbard model}
\label{sec:results}

A simple argument \cite{Scalapino1996} for an isolated plaquette already shows 
how local antiferromagnetic correlations favour $d$-wave 
pairing correlations, namely the operator that connects the antiferromagnetic $4$-particle 
ground state $|4 \rangle$ to the two-hole ground state $|2\rangle$
must have $d_{x^2-y^2}$ symmetry. The matrix element  
\begin{equation}
    \langle 2 | \Delta_{d} | 4 \rangle \ne 0 
\end{equation}
is large when the pairing operator $\Delta_{d} = (c_{3,\uparrow} c_{2,\downarrow} - c_{3,\uparrow} c_{4,\downarrow} + \cdots )$
has the sign structure for $d$-wave symmetry. 
On the other hand  $\langle 2 | \Delta_{s} | 4 \rangle = 0$ for an $s$-wave pairing operator. 
The following sections investigate the quantum state on a plaquette of the Hubbard model 
embedded in a bath of $12 \times 12$ sites which are treated numerically exactly. 

Probabilities of individual occupation number configurations are shown in Sect.~\ref{sec:diag_elements_rhoA}.
Sects.~\ref{sec:offdiag_elements_rhoA} and \ref{sec:doping_dependence} present the eigenstates 
of the plaquette reduced density matrix, with local correlations resolved according to symmetry sectors. 

\subsection{Diagonal elements of $\rho_{A=\Box}$}
\label{sec:diag_elements_rhoA}
\begin{figure}
 \centering
 \includegraphics[width=1.0\linewidth]{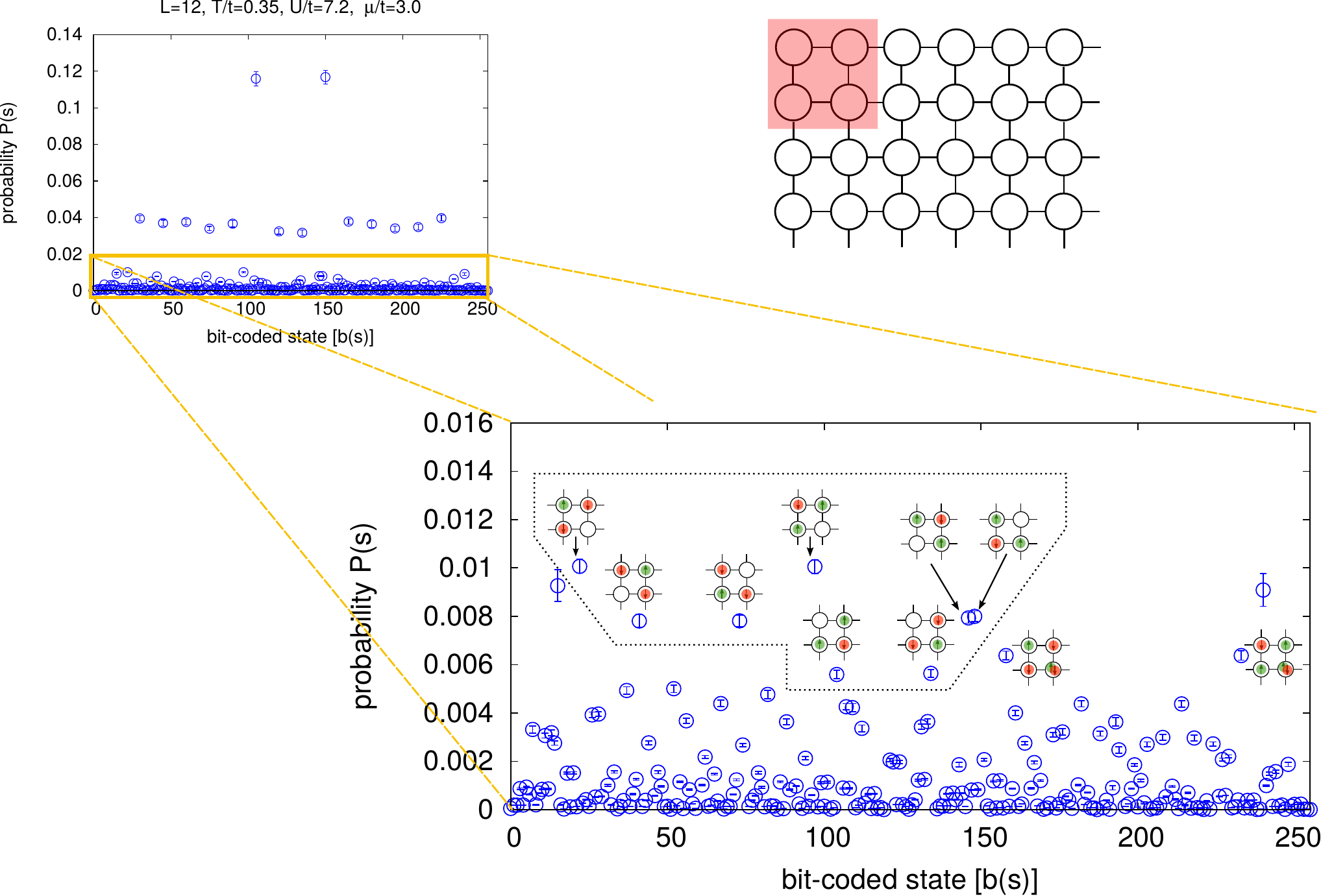}
 \caption[Probabilities of plaquette configurations in occupation number space.]
 {Probabilities $P(s)$ of particle number configurations $s$ in Fock 
 space on a plaquette that is located at the 
 corner of a large $12 \times 12$ system with open boundary conditions. 
 The binary representation of the integer $s$ encodes the Fock space configuration.
 {\bf Upper left panel:} The most probable states are the two N\'{e}el states,
 followed by the remaining 12 spin-only states.
 {\bf Main panel:} Plaquette states with charge fluctuations.
 The asymmetric location of the plaquette leads to a 
 disruption of symmetries that would be present in a translationally invariant 
 system, which is clearly visible in the probabilities. Symmetry-related states have the same probability. 
 $T/t=0.35, U/t=7.2, \mu/t=3.0 $.}
 \label{fig:plaquette_corner}
\end{figure}

Fig.~\ref{fig:plaquette_corner} shows the probabilities $P(s)$ of all plaquette configurations $s$ on
a plaquette which is located at the corner of a system with open boundary conditions.
The integer $s \in \{0,1,\ldots,255\}$ encodes the Fock configuration on a plaquette 
through its binary representation 
$[b(s)] = [n_{4}^{\uparrow} n_{3}^{\uparrow} n_{2}^{\uparrow} n_{1}^{\uparrow}n_{4}^{\downarrow} n_{3}^{\downarrow} n_{2}^{\downarrow} n_{1}^{\downarrow}]$
where $n_{i}^{\sigma}$ is the occupation number for spin $\sigma$ at one of the four sites
(shown in the upper right inset)
and the notation \verb+[ ]+ converts integer codes into bit representations.

The most probable states, the states with integer code \verb+[105]+
and \verb+[150]+, are the two N\'{e}el states; they are followed 
by the 12 other spin-only states which together would span the Hilbert 
space in a Heisenberg-like description.
From the upper left panel of Fig.~\ref{fig:plaquette_corner} one can see
that for $U=7.2$, $T/t = 0.35$ and 
half filling, on a plaquette there are 
$\sim 24 \%$  N\'{e}el states, $\sim 40 \%$ spin-only states 
(excluding the two N\'{e}el states), 
the remaining $\sim 36 \%$ are states with charge fluctuations. 

The arrangement with the plaquette at the corner, shown in the upper right inset 
in Fig.~\ref{fig:plaquette_corner}, does not possess the full symmetry of the square, the only 
symmetry operations which respect Eq.~\eqref{eq:joint_symmetry_AB}
being $\{E, C_{2x\bar{y}}\}$. This is reflected in asymmetries of the probabilities 
for plaquette configurations with a single hole (see main panel of Fig.~\ref{fig:plaquette_corner}
with plaquette configurations drawn next to representative data points):
In the presence of a boundary the hole prefers to have many neighbours 
rather than sit at the boundary which would limit the number of possible hopping processes. 
Therefore, among the configurations shown, the one with the hole located precisely at the corner 
has the lowest probability. 
Note that configurations that are related by the symmetry operation $\{E, C_{2x\bar{y}}\}$ do 
occur with the same probability. 

\begin{figure}[h!]
\centering 
\includegraphics[width=1.0\linewidth]{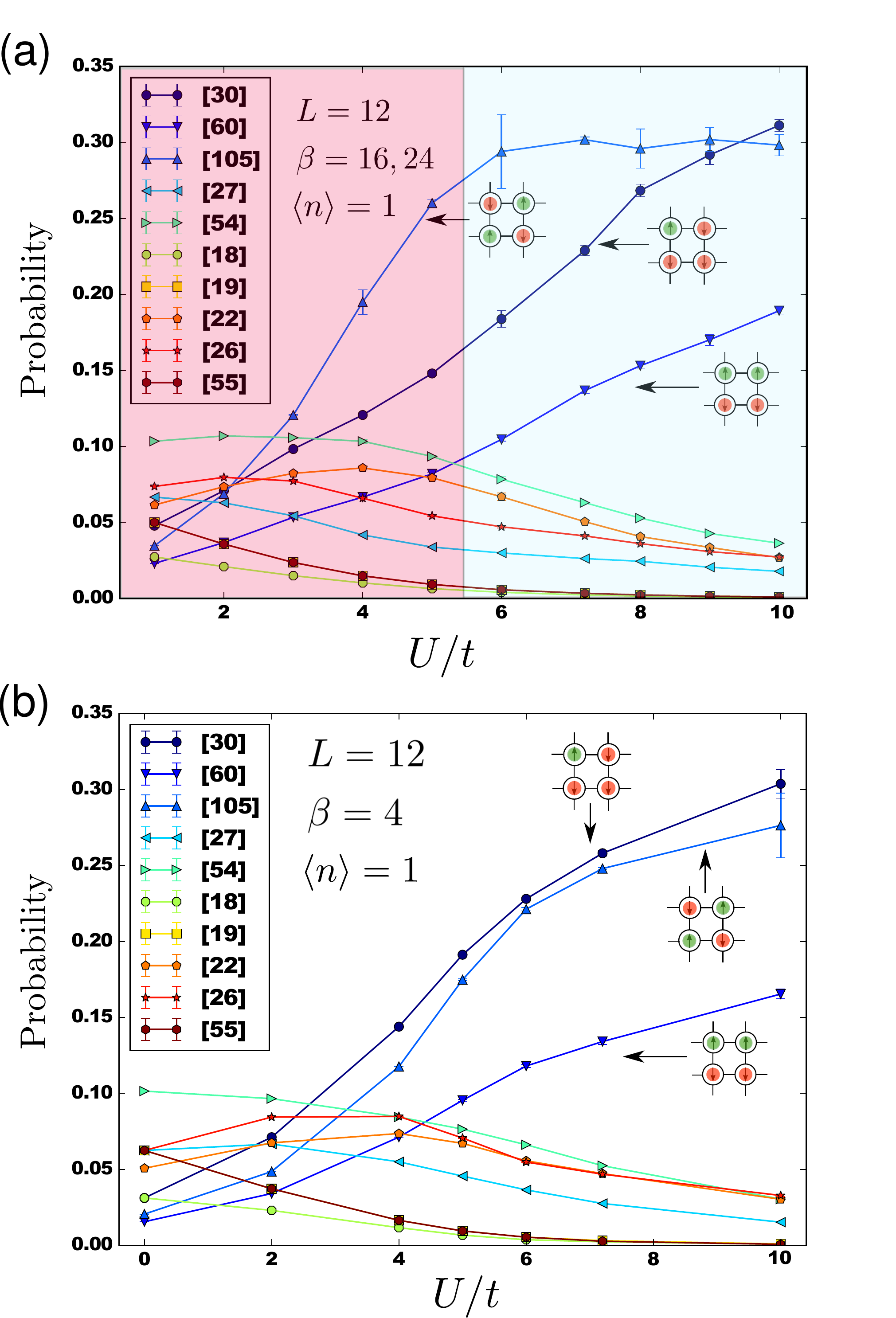}
\caption[Probabilities of selected plaquette configurations 
as a function of interaction at half filling.]
{Probabilities of selected plaquette configurations 
as a function of interaction (at half filling). The numbers in angular brackets \texttt{[ ]}
denote the bitcoded representative of a class of symmetry related plaquette configurations
(see Appendix \ref{app:plaq_symmrelated}). 
The probability shown for a particular representative is the sum of probabilities of all confgurations
in the corresponding class.}
\label{fig:plaquette_prob_vs_U}
\end{figure}

Fig.~\ref{fig:plaquette_prob_vs_U} shows the probabilities of selected plaquette occupation number states
in the repulsive Hubbard model at half filling for low temperature $\beta t = 16$ or $24$ (a)
and high temperature $\beta t = 4$ (b) as a function of the interaction strength $U/t$. 
Here, periodic boundary conditions are used so that 
the full symmetry of the square is preserved. 
Occupation number states that are related by symmetries 
are grouped into classes of states, which 
are labelled by the bitcode $[b(s)]$ of the member with the smallest bitcode 
within the class.
A list of all 34 classes of symmetry-related states with the bitcodes 
of their representatives can be found in Appendix \ref{app:plaq_symmrelated}.

At low temperature [Fig.~\ref{fig:plaquette_prob_vs_U}(a)], two datasets 
for different temperatures, {$\beta t =24$ for $U/t \le 5$ and 
$\beta t = 16$ for $U/t \ge 6$, have been combined, which is indicated by 
different colours of the background shading. 
For $U/t \ge 3$, the most probable states are the two N\'{e}el 
states. The second most important class of states \verb+[30]+ comprises 
those states with three particles of one spin and one particle of the opposite 
spin. Taking into account spin-inversion symmetry 
there are 8 such states and their probabilities have been 
summed up in Fig.~\ref{fig:plaquette_prob_vs_U}. This explains why this class 
of states appears with higher probability in Fig.~\ref{fig:plaquette_prob_vs_U}
than the class of states \verb+[60]+, consisting of only 4 states,
namely the spin configurations that are FM aligned in one direction and 
AFM aligned in the other one. The multiplicities of different classes of states,
which are essential for interpreting Fig.~\ref{fig:plaquette_prob_vs_U}, 
are also given in Appendix \ref{app:plaq_symmrelated}.

To illustrate that detailed information can be gleaned from 
configurational probabilities we consider next in Fig.~\ref{fig:plaquette_prob_vs_U}
the classes \verb+[27]+ and \verb+[54]+, which 
both have a multiplicity of 16 states. 
The class of states \verb+[27]+, 
comprising plaquette configurations 
with a neighbouring doublon-hole pair 
and the singly-occupied sites in a FM configuration, 
has smaller probability than states \verb+[54]+ 
representing a neighbouring doublon-hole pair
with the singly-occupied sites arranged in an AFM configuration.
Note also that the classes of states \verb+[19]+ and \verb+[55]+ have exactly 
the same probability due to particle-hole symmetry at half filling. 

The plaquette probabilities at high temperature $\beta t = 4$
[Fig.~\ref{fig:plaquette_prob_vs_U}(b)]
are qualitatively very similar to those at low temperature,
indicating that local correlations of the low-temperature phase 
are already well developed at $\beta t = 4$.

In Fig.~\ref{fig:plaquette_prob_vs_n}(a) the doping dependence of the plaquette
probabilities is presented for experimentally relevant inverse temperature 
$\beta t=4$ and for repulsive interaction $U/t=2, 4,$ and $7.2$.
Error bars
in Fig.~\ref{fig:plaquette_prob_vs_n}(a) are deduced from the spread 
of datapoints within one class of states which should have 
the same probability due to symmetry.
\begin{figure}[h!]
 \centering 
 \includegraphics[width=1.0\linewidth]{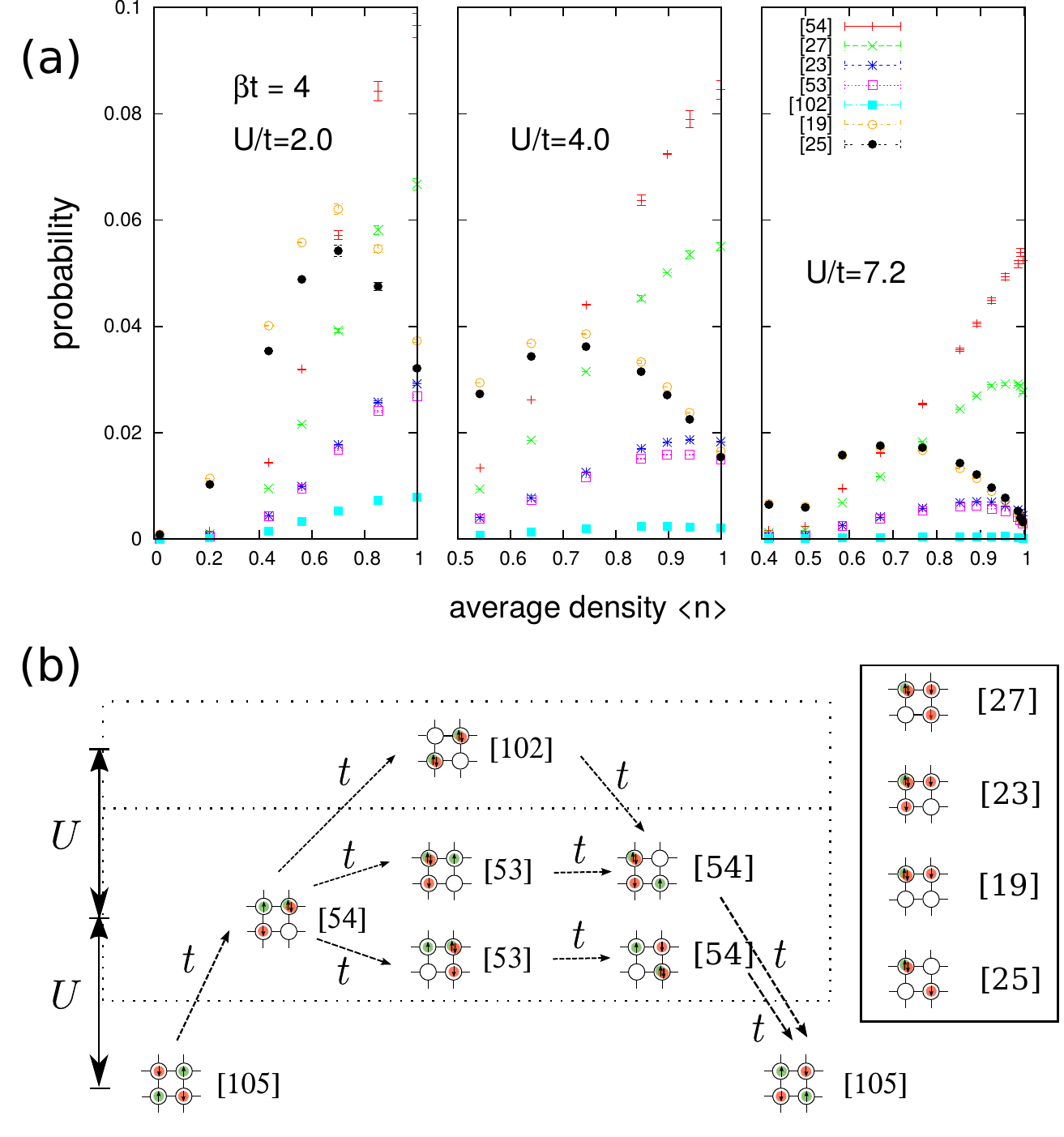}
 \caption[Doping dependence of the probabilities of selected plaquette configurations.]
 {(a) Doping dependence of the probability of selected plaquette configurations with at least one doubly occupied site;
 $\beta t = 4$, $U/t=2,4,7.2$. 
 The total system size is $L \times L$ with $L=12$. 
 The aggregated probabilities are the sum over all states that are related by lattice or 
 spin inversion symmetry (see Appendix \ref{app:plaq_symmrelated}).
 Note the non-monotonic behaviour at quarter filling $\langle n \rangle = 0.5$
 in subfigure (a) for $U/t=7.2$.
 (b) Typical pathways of ring exchange processes in fourth
 order perturbation theory, where some of the plaquette configurations in (a) appear as intermediate states. Away from half filling also 
 third order spin exchange processes with intermediate states such as \texttt{[19]} or \texttt{[25]} are present.
 }
 \label{fig:plaquette_prob_vs_n}
\end{figure}
The selected plaquette configurations have at least one doubly occupied 
site so that for large Hubbard repulsion they represent the intermediate 
virtual states through which pairwise and ring exchange interactions 
in an effective 
spin Hamiltonian are mediated \cite{Takahashi1977,MacDonald1988,Delannoy2005}. 
A typical pathway of hopping processes leading in fourth 
order perturbation theory to ring exchange 
interactions is illustrated in Fig.~\ref{fig:plaquette_prob_vs_n}(b). 

The overall trend is that the amplitude of states with charge 
fluctuations is reduced with increasing Hubbard repulsion. 
The amplitude of states with neighbouring doublon-hole pairs 
decreases as $t/U$, in accordance with second order perturbation theory. 
Plaquette configurations with a doublon-hole pair on diagonally
opposite corners (\verb+[53]+),
occurring  as intermediate states in ring-exchange pathways [Fig.~\ref{fig:plaquette_prob_vs_n}(b)],
have a very small probability at all fillings, which is for large $U/t$
approximately an order of magnitude smaller 
than that of states with neighbouring doublon-hole 
pairs (\verb+[27]+,\verb+[54]+), as to be expected on the basis 
of fourth order perturbation theory.

Based on Fig.~\ref{fig:plaquette_prob_vs_n}(a),
very detailed observations regarding correlation effects can be made.
For example, a signal for local correlation lies in the differences 
of probability between similar configurations such as \verb+[23]+ and \verb+[53]+
or \verb+[19]+ and \verb+[25]+,
which would be equally likely, if the singly and doubly occupied sites 
were placed on the lattice randomly, with say probability $p_d$
for a doubly occupied site and $p_s$ for a singly occupied one.  
It is important to note that \verb+[19]+ has multiplicity 16, 
whereas \verb+[25]+ has 
multiplicity 8 (see Appendix \ref{app:plaq_symmrelated}). 
Thus, the observation that the classes \verb+[19]+ and \verb+[25]+ 
have almost the same probabilities in
Fig.~\ref{fig:plaquette_prob_vs_n}(a)
indicates that the probability per individual configuration 
of states \verb+[25]+ is kinetically 
enhanced compared to states from \verb+[19]+, since the former
allow for more hopping processes on the plaquette. 
Note also in Fig.~\ref{fig:plaquette_prob_vs_n}(a) at $U/t=7.2$
the discontinuous jump at quarter filling $\langle n \rangle \approx 0.5$.


\subsection{Off-diagonal elements of $\rho_{A=\Box}$}
\label{sec:offdiag_elements_rhoA}
With the knowledge of all off-diagonal elements of the reduced density matrix, we can compute 
the entanglement spectrum and resolve it according to symmetry sectors. 
\begin{figure}[b!]
\centering
\includegraphics[width=1.0\linewidth]{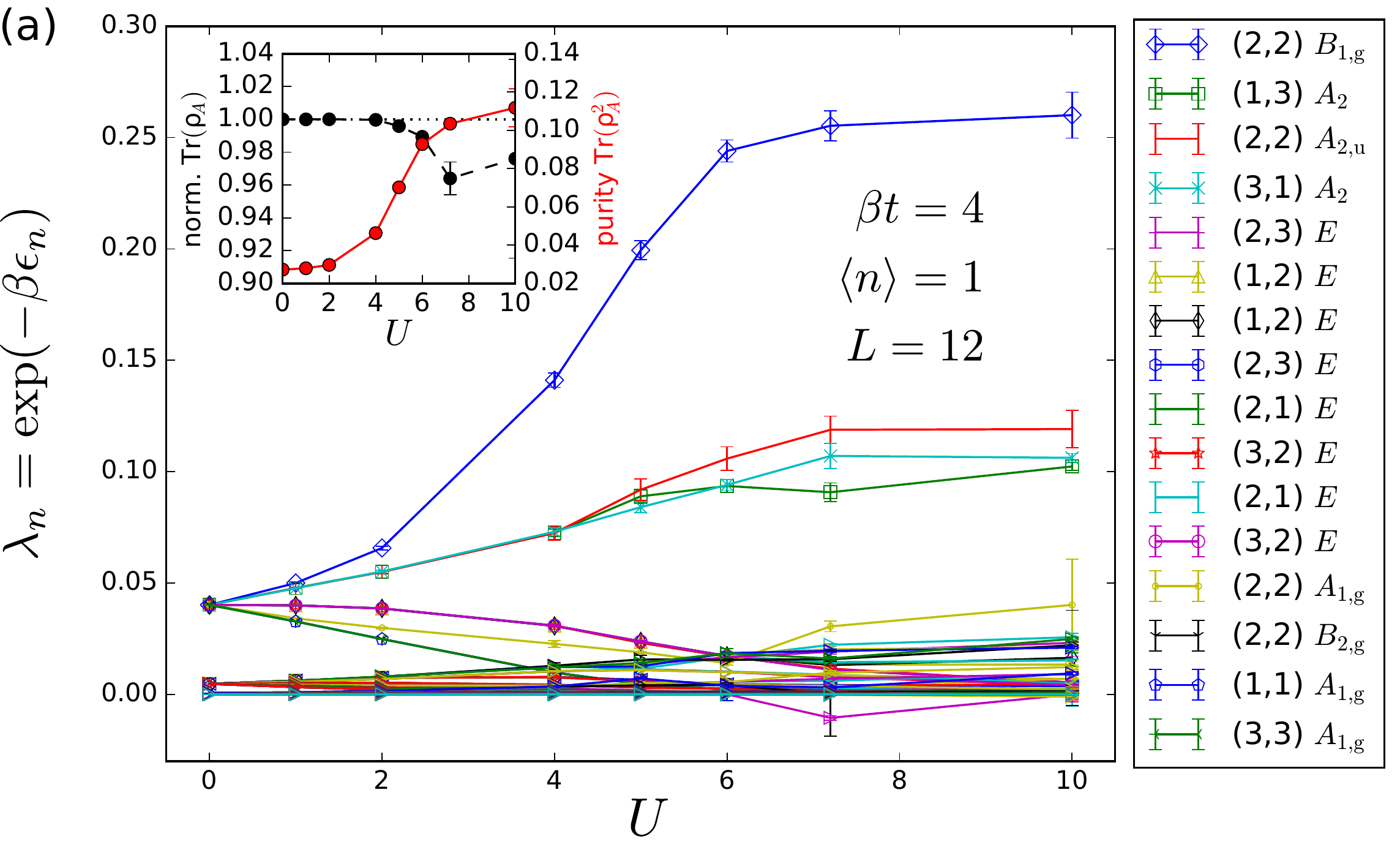}
\includegraphics[width=1.0\linewidth]{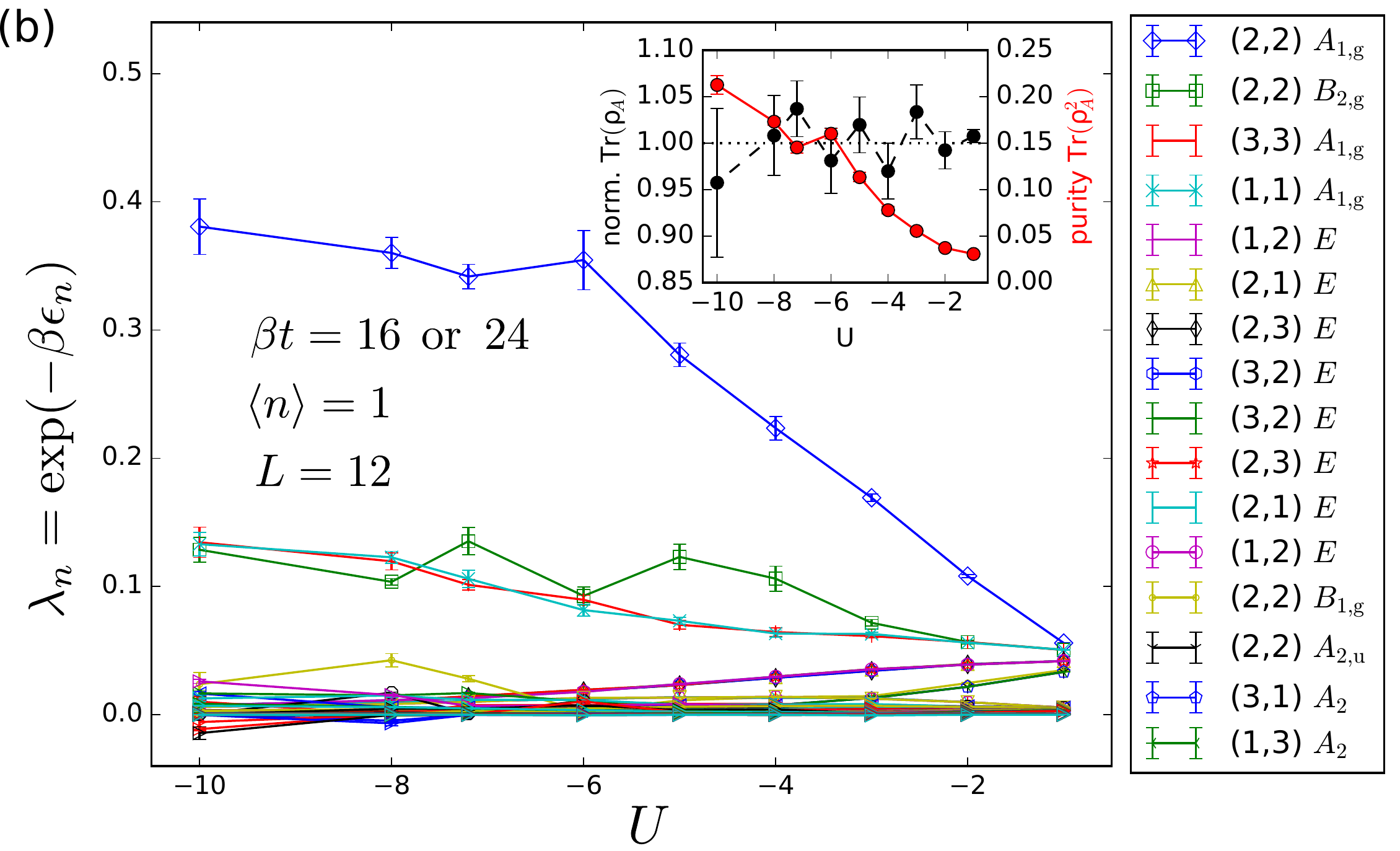}
\caption[Eigenvalues of the reduced density matrix at half filling, labelled according to symmetry sectors.]
{Eigenvalues $\{\lambda_{n}\}_{n=1}^{4^{N_s}=256}$ of the reduced density matrix $\rho_{A=\Box}$ on a plaquette
at half filling and inverse temperature (a) $\beta t = 4$ and (b) $\beta t = 24$ (for $|U|/t \le 5$) or $\beta t = 16$ (for $|U|/t \ge 6 $).
Eigenvalues are labelled according to irreducible representations of the symmetry group $D_4$ of the square
as well as spin inversion symmetry. The plaquette is embedded in an 
$L \times L$ system with $L=12$ and periodic boundary conditions. 
The inset shows the normalization $\text{Tr}\left(\rho_{A=\Box}\right)$ and purity $\text{Tr}\left(\rho_{A=\Box}^2\right)$
of the reduced density matrix.}
\label{fig:eigvals_rhoA_allU}
\end{figure}

Fig.~\ref{fig:eigvals_rhoA_allU} shows the eigenvalue spectrum 
of the plaquette reduced density matrix $\rho_{A=\Box}$
at half filling as a function 
of Hubbard interaction for high (a) and low (b) temperature.
The 16 most important eigenstates 
of $\rho_{A=\Box}$ are labelled by their symmetry sectors $(N_{A,\uparrow}, N_{A,\downarrow}) M$
where $(N_{A,\uparrow}, N_{A,\downarrow})$ is the particle number sector and $M$ is the Mulliken symbol 
describing the symmetry of the state under the operations of the symmetry group $D_{4h}$ 
which combines the symmetries of the square and spin inversion symmetry. 
The data for high temperature in Fig.~\ref{fig:eigvals_rhoA_allU}(a) is presented for the 
repulsive Hubbard model, while for the low temperature data in Fig.~\ref{fig:eigvals_rhoA_allU}(b) we have 
chosen the language of the attractive Hubbard model. At half filling, both models are exactly equivalent;
the eigenstates of $\rho_{A=\Box}$ for the repulsive Hubbard model are related to those 
of the attractive model by the spin-down particle-hole transformation
\begin{equation}
 c_{i,\downarrow} \longrightarrow (-1)^{\vecbf{i}} c_{i,\downarrow}^{\dagger}, \quad c_{i,\uparrow} \longrightarrow c_{i,\uparrow},
 \label{eq:Shiba_trafo}
\end{equation}
where $(-1)^{\vecbf{i}} \equiv (-1)^{i_x+i_y}$ is a staggered phase factor for one of the two sublattices of the square lattice. 
We point out that equivalent states in both models behave differently under symmetry operations 
of the point group, which is reflected in their Mulliken symbols. 
There is a one-to-one correspondence between the Mulliken symbols of equivalent states in the repulsive 
and attractive model, for example states with symmetry $(2,2) A_{2,u}$ in the language of the positive-$U$ model
correspond to states with symmetry $(2,2) B_{2,g}$ in the negative-$U$ model, etc. 
We stress that in the presentation of Fig.~\ref{fig:eigvals_rhoA_allU}(a) and (b) same colours do not necessarily imply that 
the states are equivalent in the two models~\footnote{  
The colours are based on the sequence of energy levels at $|U|/t=4$.}.

The insets in Fig.~\ref{fig:eigvals_rhoA_allU}(a) and (b) show the normalization $\text{Tr}\left(\rho_{A=\Box} \right)$
and the purity  $\text{Tr}\left(\rho_{A=\Box}^2 \right)$ of the plaquette reduced density matrix. 
The purity increases with interaction strength, signalling that the state becomes closer to the $T=0$ limiting case of 
a product state of non-entangled plaquettes. For non-interacting fermions ($U=0$) the plaquette is most mixed
with its environment; then there is a pronounced upturn in $\text{Tr}\left(\rho_{A=\Box}^2 \right)$ around $U=5$,
which levels off for $U \rightarrow 10$. The purity at $\beta t = 4$ (Fig.~\ref{fig:eigvals_rhoA_allU}(a)) 
is by a factor of two smaller than at $\beta t = 24$ (Fig.~\ref{fig:eigvals_rhoA_allU}(b)), 
as to be expected due to thermal entropy. $\rho_{A=\Box}$ is normalized 
within error bars, which, however, increase for larger $U$ and lower temperature. 

Focusing first on results for high temperature Fig.~\ref{fig:eigvals_rhoA_allU}(a), we observe that 
at $U=0$ the 16 most important eigenstates are all degenerate and clearly 
separated from the remaining eigenstates with lower weight. As the interactions are switched on, 
the multiplet splits into two singlet-triplet type sequences, namely the four states 
$|s^{(1)}\rangle \in (2,2) B_{1,g}$ and 
\mbox{$\left\{|t^{(1)}_{0}\rangle \in (2,2) A_{2,u},\,
|t^{(1)}_{-}\rangle \in (1,3) A_2,\,
|t^{(1)}_{+}\rangle \in (3,1) A_2 \right\}$},
and the four states 
$|s^{(2)}\rangle \in (2,2) A_{1,g}$ and 
\mbox{$\left\{|t^{(2)}_{0}\rangle \in (2,2) B_{2,g},\,
|t^{(2)}_{-}\rangle \in (1,1) A_{1,g},\,
|t^{(2)}_{+}\rangle \in (3,3) A_{1,g} \right\}$},
and into a degenerate octet of eigenstates with $p$-wave symmetry (labelled by the 
irreducible representation label $E$). 

The states with $p$-wave symmetry are both spin 
and pseudospin doublets, which together with the fact that their
irreducible representation $E$ is two-dimensional 
explains their eightfold degeneracy \cite{Schumann2002}.
The degeneracy will be lifted either by applying an external magnetic field 
or by shifting the chemical potential away from the half filling point $\mu=\frac{U}{2}$.
It is remarkable how well the degeneracy of the octet is preserved in the Monte Carlo data 
of Fig.~\ref{fig:eigvals_rhoA_allU}(a) and (b).

In comparison with the grand canonical eigensystem of an isolated
4-site Hubbard model, which is worked out analytically in Ref.~\cite{Schumann2002},
the succession of energy levels
appears changed in  Fig.~\ref{fig:eigvals_rhoA_allU}(a) and (b)
in that the ``entanglement energy'' of the octet is lower (i.e.
it has a heigher weight in the thermal state)
than that of the low-weight singlet-triplet-structure  $\{|s^{(2)}\rangle, |t^{(2)}_{\pm,0}\rangle\}$ for $|U|<6$.
This shows that the method presented here can resolve fine differences between the energy spectrum 
of an isolated plaquette \cite{Schumann2002}
and of a plaquette embedded in a much larger system. 
Note that it is not simply the difference in temperature between the ground 
state spectrum of Ref.~\cite{Schumann2002}
and our ``entanglement energy`` spectrum, which could explain this discrepancy,
since it occurs also at $\beta = 24$ in Fig.~\ref{fig:eigvals_rhoA_allU}(b) and 
since temperature cannot change the relative order of the statistical weights.

For $U>4$, the degeneracy of the high-weight triplet $|t^{(1)}_{\pm,0}\rangle $ is lifted,
and concomitantly the purity of the reduced density matrix increases 
(see inset Fig.~\ref{fig:eigvals_rhoA_allU} (a)). The fact that the octet
of eigenstates remains perfectly degenerate in the interval $4 \le U \le 6$ 
supports the picture that the lifting of the degeneracy in $|t^{(1)}_{\pm,0}\rangle$ is not merely an artifact 
of larger error bars. 
The low-weight singlet state $|s^{(2)} \rangle \in (2,2) A_{1,g}$ shows non-monotonic behaviour as 
a function of $U$ and its weight appears to increase again for $U > 6$. However, the error 
bars are too large to draw any conclusions. 

In Fig.~\ref{fig:eigvals_rhoA_allU}(b) entanglement spectra for two low temperatures,
$\beta t = 24$ (for $|U|/t \le 5$) and $\beta t = 16$ (for $|U|/t \ge 6$), are combined.
Due to issues of ergodicity at large Hubbard interactions it was not possible to reach 
lower temperatures for $|U|/t \ge 6$.
The eigenvalue spectra $\{\lambda_n\}$ have a qualitatively similar dependence on $U$
both for high and low temperature. 
In both cases, multiplets of degenerate states that exist in the sector with charge fluctuations for small $U$
mix for large $|U|/t$ and give rise to a broad structureless ''band`` of small eigenvalues, from 
which a low-lying singlet state seems to separate off. However, large error bars prevent a 
conclusive statement. 

When discussing now to the low-temperature spectrum 
of $\rho_{A=\Box}$ displayed in Fig.~\ref{fig:eigvals_rhoA_allU}(b), we refer to the Mulliken symbols
shown next to Fig.~\ref{fig:eigvals_rhoA_allU}(b). 
It must be stessed again that equivalent states in the repulsive and attractive Hubbard model, 
i.e. states related by the spin-down particle-hole transformation Eq.~\eqref{eq:Shiba_trafo}, 
are labelled by different Mulliken term symbols. 
For the positive-$U$ Hubbard model,
the singlet-triplet type structures are comprised of the four states
 $|s^{(1)} \rangle \in (2,2)A_{1,g}$ and 
 $\{|t_{0}^{(1)} \rangle \in (2,2) B_{2,g}, \,|t_{+}^{(1)} \rangle \in (3,3) A_{1,g}, \,|t_{-}^{(1)} \rangle \in (1,1) A_{1,g}\}$
 and the four states
 $|s^{(2)} \rangle \in (2,2)B_{1,g}$ and 
 $\{|t_{0}^{(2)} \rangle \in (2,2)A_{2,u}, \,|t_{+}^{(2)} \rangle \in (3,1)A_2, \,|t_{-}^{(2)} \rangle \in (1,3)A_2\}$.
 In between the two singlet-triplet structures there is again an octet of degenerate states. 
A remarkable difference to the high-temperature spectrum is that the degeneracy of the high-weight triplet 
is lifted at smaller $|U|/t$, namely at $|U|/t=2$ for $\beta t=24$ compared to $U/t=4$ for $\beta t=4$.
The weight of $|t_0^{(1)} \rangle $ increases rapidly, the subsequent decrease at $|U|/t=6$ must be attributed 
to the change in temperature from $\beta t=24$ to $\beta t=16$ when changing from the dataset with $\beta t=24$ for $|U|/t \le 5$
to the dataset with $\beta t = 16$ for $|U|/t \ge 6$.

\subsection{Doping dependence}
\label{sec:doping_dependence}
Next, we turn to the doping dependence of the plaquette entanglement spectrum, which 
is displayed in Fig.~\ref{fig:eigvals_rhoA_beta4U6_mu} for the repulsive Hubbard model at $U/t=6$ and 
high temperature $\beta t = 4$.
The DQMC algorithm suffers 
from a sign problem \cite{Loh1992} in the repulsive Hubbard model when particle-hole symmetry is broken 
by tuning the chemical potential away from the half-filling point $\mu = \frac{U}{2}$.
Nevertheless, at relatively high temperature around $\beta t=4$ 
simulations are still possible due to a mild sign problem \cite{Iglovikov2015}
that can be offset by acquiring more statistics in longer Monte Carlo runs. 
Fig.~\ref{fig:eigvals_rhoA_beta4U6_mu} shows the 
eigenvalue spectrum $\{\lambda_n\}_{n=1}^{4^{N_s}}$ of the plaquette $(N_s=4)$ reduced 
density matrix $\rho_{A=\Box}$ with coloured stripes indicating 
blocks of fixed particle number $(N_{\uparrow}, N_{\downarrow})$.

\begin{figure*}[t!]
\centering 
\includegraphics[width=1.0\textwidth]{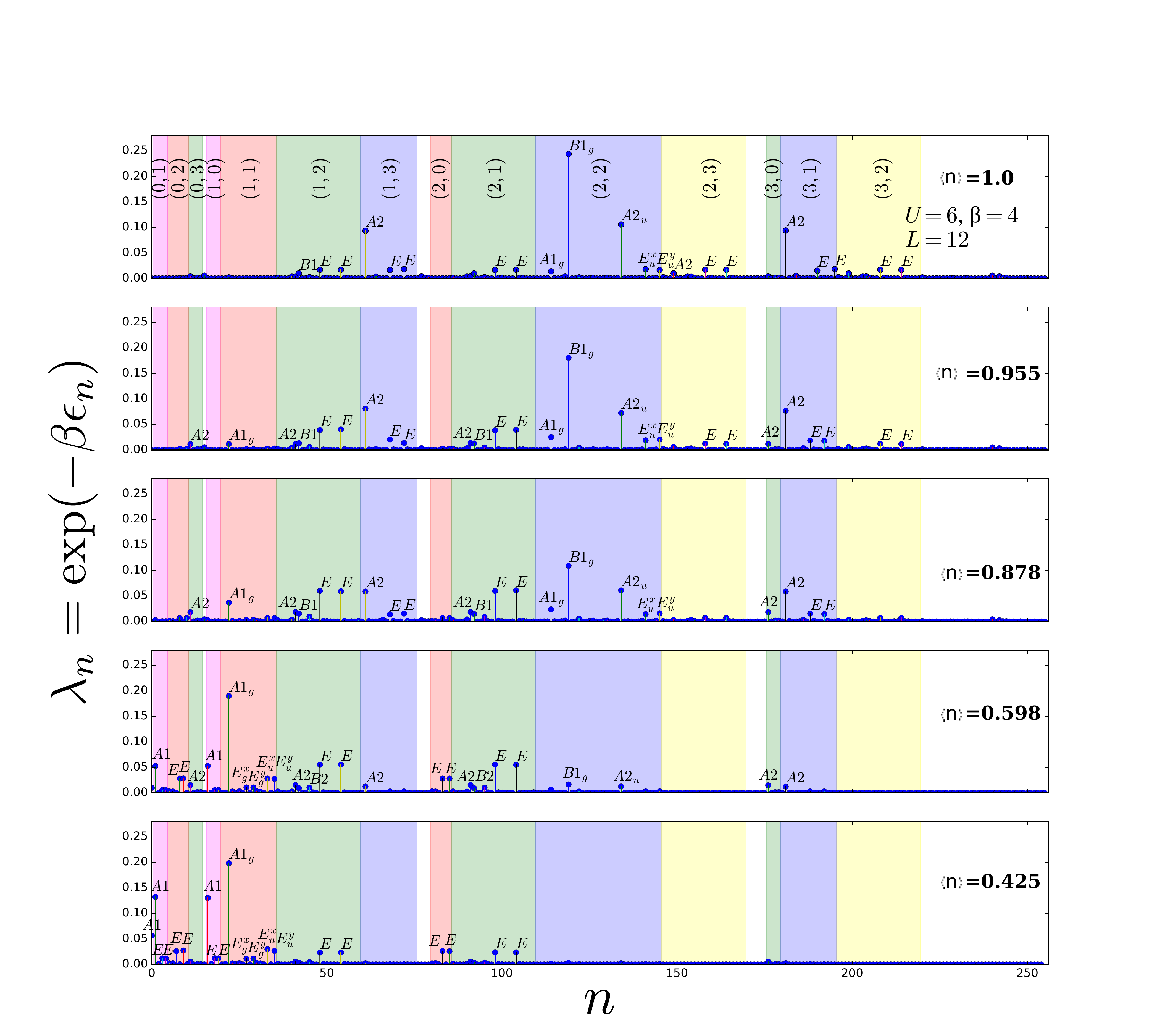}
\caption[Eigenvalues of the reduced density matrix as a function of doping.]
{Eigenvalue spectrum $\{\lambda_n\}_{n=1}^{4^{N_s}}$ of the reduced density matrix $\rho_{A=\Box}$
for a plaquette ($N_s=4$), organized into particle number sectors, which are indicated by coloured segments. 
$U/t=6$, $\beta t = 4$, and total linear system size $L=12$; rows from top to bottom correspond
to decreasing filling $\langle n \rangle$.
For clarity, only eigenvalues larger than $0.01$ are labelled by the irreducible 
subspace to which the corresponding eigenvector belongs.
Error bars (not shown) are of the order of $1-5\%$.}
\label{fig:eigvals_rhoA_beta4U6_mu}
\end{figure*}

In the following, the most important eigenstates from the respective symmetry multiplets 
are listed explicitly. 
At half filling, the leading eigenstate of the plaquette reduced density matrix has $d_{x^2-y^2}$-wave ($B_{1g}$) symmetry,
whereas around quarter filling, $\langle n \rangle \approx 0.5$, it
has $s$-wave ($A_{1g}$) symmetry (see first and fourth row of Fig.~\ref{fig:eigvals_rhoA_beta4U6_mu}). 
The invariant subspace labelled by  $(2,2) B_{1g}$ consists of four states and 
$(2,2) A_{2u}$ consists of three states
(see Tab.~\ref{tab:group_structure_invsym}),
but it turns out that a single state from each symmetry multiplet has by far the largest coefficient,
namely:
\begin{subequations}
\begin{align}
 &(2,2)B_{1,g} \ni |\psi_1 \rangle   \sim \frac{1}{\sqrt{2}} (|\downarrow \, \uparrow \, \uparrow \, \downarrow \rangle + |\uparrow \, \downarrow \, \downarrow \, \uparrow \rangle ) + \ldots \label{eq:eigenstate_22B1g_largest_coeff} \\ 
 &(2,2)A_{2,u} \ni |\psi_2 \rangle   \sim \frac{1}{\sqrt{2}}(| \downarrow \uparrow \uparrow \downarrow \rangle - | \uparrow \downarrow \downarrow \uparrow \rangle ) + \mathcal{O}(t/U). \label{eq:eigenstate_22Au_largest_coeff}
\end{align}
\end{subequations}
Dots in Eq.~\eqref{eq:eigenstate_22B1g_largest_coeff} indicate states from the same 
symmetry multiplet without double occupancy but with much smaller weight. 
Thus, the leading eigenvectors of $\rho_{A=\Box}$ in the particle number sector $(2,2)$ at half filling are the symmetric and antisymmetric combinations of 
the two N\'{e}el states, as to be expected. 
Note that for $U \gtrsim 6$ an additional $p$-wave doublet $(E_u^{x}, E_{u}^{y})$ 
appears in the particle number sector $(2,2)$
(see first row of Fig.~\ref{fig:eigvals_rhoA_beta4U6_mu}).
The leading eigenvectors in the particle number sector $(1,1)$ at quarter filling are 
\begin{equation}
\begin{split}
  (1,1) A_{1,g} \ni | \psi_3 \rangle  \sim \frac{1}{2} ( | \uparrow \, h \,h\,\downarrow \rangle + | h \, \uparrow \, \downarrow \, h \rangle  \\
					      + | h\, \downarrow \, \uparrow \, h \rangle + | \downarrow \, h \,h \, \uparrow \rangle ) + \ldots,
\end{split}					      
\label{eq:eigenstate_11A1g_largest_coeff}					      
\end{equation}
i.e. two holes in diagonally opposite corners, which maximizes their kinetic energy,
and a set of states with smaller weight with two neighbouring holes, which have $p$-wave symmetry
(labelled by $E_{u(g)}^{x(y)}$ in the fourth row in Fig.~\ref{fig:eigvals_rhoA_beta4U6_mu}).
The character of the leading eigenstates in the particle number sectors $(2,2)$ and $(1,1)$ 
as described by Eqs.~\eqref{eq:eigenstate_22B1g_largest_coeff} 
and \eqref{eq:eigenstate_11A1g_largest_coeff} hardly changes with doping. 
The leading eigenstate from the symmetry sector $(1,3)A_{2}$
\begin{equation}
\begin{split}
 (1,3)A_{2} \ni |\psi_4 \rangle \sim \frac{1}{2}  ( | \uparrow \, \downarrow \, \downarrow \, \downarrow \rangle + | \downarrow \, \uparrow \, \downarrow \, \downarrow \rangle \\
                                                    -| \downarrow \, \downarrow \, \uparrow \, \downarrow \rangle - | \downarrow \, \downarrow \, \downarrow \, \uparrow \rangle 
                                                  )
                                              + \mathcal{O}(t/U)   
\end{split}                                              
\label{eq:eigenstate_31A2_largest_coeff}                                              
\end{equation}
and its spin-reversed counterpart from the symmetry sector $(3,1)A_{2}$ are degenerate with 
the leading eigenstate from $(2,2)A_{2,u}$ (Eq.~\eqref{eq:eigenstate_22Au_largest_coeff})
up to $U/t \lesssim 4$ (at $\beta t = 4$).
Finally, we list the state 
\begin{equation}
\begin{split}
 (2,2) A_{1,g} \ni |\psi_5 \rangle \sim \frac{1}{2}   ( |\uparrow \, \uparrow \, \downarrow \, \downarrow \rangle + |\uparrow \, \downarrow \, \uparrow \, \downarrow \rangle  \\
							   +|\downarrow \, \uparrow \, \downarrow \, \uparrow \rangle + | \downarrow \, \downarrow \, \uparrow \, \uparrow \rangle 
                                                      ) + \mathcal{O}(t/U),
\end{split}                                                    
\end{equation}
with spins that are ferromagnetically aligned along one coordinate axis and antiferromagnetically in the other direction and 
which separates off from the ``band'' of low-lying states for large $|U|$ (see Fig.~\ref{fig:eigvals_rhoA_allU}(a)).

\section{Conclusion and outlook} 
\label{sec:conclusion}
We have provided proof of principle calculations
that it is possible in an equilibrium DQMC simulation
to obtain the full reduced density matrix of a small subsystem embedded
in a much larger system that can be interpreted as the exact correlated bath. 
The sequence of ``entanglement energy'' levels of the embedded plaquette 
is shown to differ from the sequence of levels for an isolated plaquette \cite{Schumann2002}. 

The calculated configurational probabilities allow detailed 
benchmarking of current fermionic quantum gas microscope experiments.

The possiblity of computing the full quantum state of a subsystem is 
a unique feature of the DQMC framework, which is based on the free fermion 
decomposition \cite{Grover2013}.
Due to the factorization of the Monte Carlo weight into a spin-$\uparrow$
and spin-$\downarrow$ part, the computational cost for obtaining all elements of $\rho_A$
scales like $2\times (2^{N_s} \times 2^{N_s})$, rather than $4^{N_s} \times 4^{N_s}$, 
in one Monte Carlo sample. 
However, the storage requirement for all elements is $4^{N_s} \times 4^{N_s}$,
which is forbidding for e.g. $N_s=9$.
If individual (particle number and point group) symmetry sectors are targeted 
by performing the transformation Eq.~\eqref{eq:occ_to_rep} in every Monte Carlo step
rather than computing all elements of $\rho_A$ in the occupation number basis, 
the number of non-vanishing matrix 
elements of $\rho_A$ that need to be kept for Monte Carlo averaging 
can be reduced to $a_n \times a_n$, where $a_n$
is the number of copies of the $n$-th irreducible representation that appear 
in the decomposition of a given particle number sector (see Eq.~\eqref{eq:n_occur_irrep}).
The additional cost of the basis transformation Eq.~\eqref{eq:occ_to_rep} in every 
Monte Carlo step can be compensated by a finer granularity of the parallelization.
This would give access to $3 \times 3$ subsystems that 
can already capture the effect of 
next nearest neighbour hopping $t^{\prime}$, which needs to be included 
to describe qualitatively the electronic band structure of cuprates. 
At least for temperatures, where DQMC
simulations are still possible in spite of
the sign problem \cite{Iglovikov2015}, one may thus hope to gain some insight 
into the role of local correlations
in the high-temperature phase of a prototypical model for high-$T_c$
superconductors, whose pseudogap regime and 
anomalous normal state commonly referred to as ``strange'' 
metal phase is still poorly understood \cite{Keimer2015}.

A model for which quantum state tomography 
on a plaquette is particularly meaningful is 
the plaquette Hubbard model studied in Refs.~\cite{Rey2009} and \cite{Ying2014},
which interpolates between isolated plaquettes and a uniform square lattice 
taking the interplaquette hopping as a tunable parameter. 
The presented approach may also prove useful for computing 
dynamical properties such as the spectral function or optical conductivity
without the need for analytical continuation of imaginary-time correlation 
functions, in some form of cluster approximation (see e.g. Ref.~\cite{Senechal2000}),
albeit with an exact correlated bath.

\section*{Acknowledgments}
The author thanks Lei Wang for helpful discussions
and for pointing out Refs.~\cite{Udagawa2010, Udagawa2015}.
The numerical simulations were performed on \mbox{JURECA}, 
J\"{u}lich Supercomputing Center. 
Support by the International Young Scientist Fellowship
of Institute of Physics, Chinese Academy of Sciences under the Grant No. 2018004
is acknowledged. 

\appendix

\section{List of symmetry-related classes of states for a single plaquette}
\label{app:plaq_symmrelated}

Each of the 256 plaquette states 
is labelled  by an integer $x$ between 0 and 255 with the convention that its binary representation
$[x]$ corresponds to the occupation numbers on the plaquette; the four least significant bits 
denote occupation numbers for spin-$\downarrow$ (see Fig.~\ref{fig:symm-related_states}).
\begin{figure}
\centering
\includegraphics[width=1.0\linewidth]{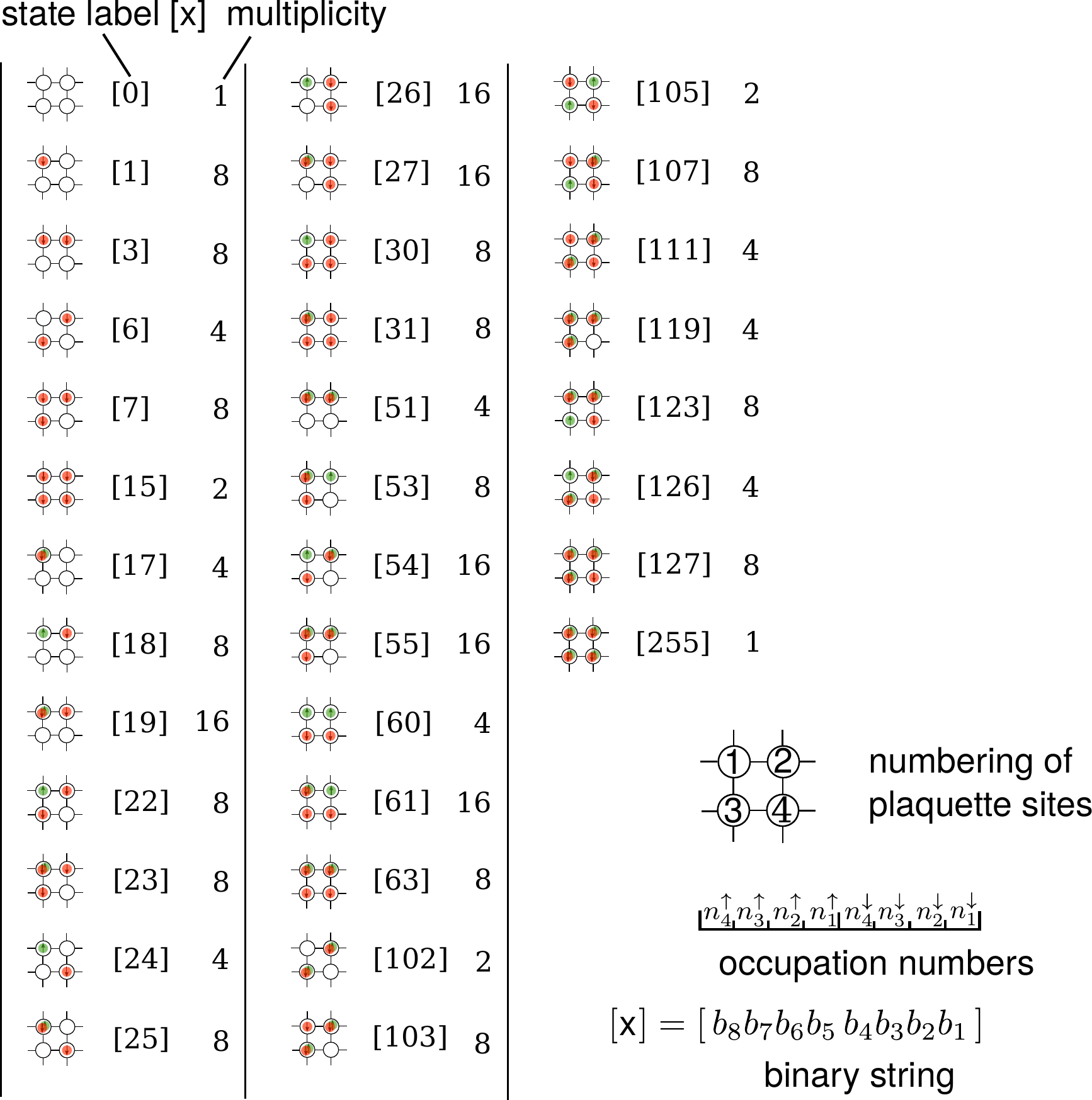}
\caption{Representative configurations for the classes of symmetry-related plaquette configurations.}
\label{fig:symm-related_states}
\end{figure}  
The leftmost column of Fig.~\ref{fig:symm-related_states} shows microscpic configurations
of spin-$\uparrow$  and spin-$\downarrow$
particles on the plaquette. States that are related by a symmetry operation of the point group 
$D_4$ of the square or by spin-inversion are grouped into classes of symmetry-related states.
The 34 classes are listed with a representative spin configuration
for each class, its bitcoded label and the number of symmetry-related states in the class (``multiplicity'').
We choose as a representative from each class the state with the smallest bitcoded label.
Special sets of states are the two N\'{e}el states (labelled by $[105]$), the 16
states in the spin-only subspace (with representatives $[15]$,$[30]$,$[60]$ and $[105]$), the 
states with neighbouring doublon-hole pairs in a spin-only background ($[27]$, $[54]$) and 
the states with diagonally-opposite doublon-hole pairs in a spin-only background ($[23]$,$[53]$).

\section{Group theoretic techniques: projection operator method}
\label{app:projection_operator_method}
The relation between transformation operators acting in the many-body Hilbert space 
and the symmetry operations acting on coordinates is provided by 
Wigner's convention \cite[Chapt.~3]{TinkhamGroupTheory}
\begin{equation}
 \hat{P}_R f(R x) = f(x) \Leftrightarrow \hat{P}_R f(x) = f(R^{-1}x),
 \label{eq:Wigner_convention}
\end{equation}
where $\hat{P}_R$ is the operator acting onto wave functions in second quantization
while the symmetry operator $R$ acts onto indices of creation and annihilation operators. 

The fermion ordering in the definition of the states is chosen such
that site indices of creation operators increase from right to left 
and creation operators for $\uparrow$-particles are to the left of operators 
for $\downarrow$-particles, e.g. 
$| \uparrow_4 \, \downarrow_3 \, \uparrow_2 \, \downarrow_1 \, \rangle  \equiv c_{4\uparrow}^{\dagger} c_{2 \uparrow}^{\dagger} c_{3 \downarrow}^{\dagger} c_{1 \downarrow}^{\dagger} | \text{vac} \rangle$.
As an illustration of Eq.~\eqref{eq:Wigner_convention} and of the action of the symmetry operators 
$\hat{P}_R$ on the many-body Hilbert space, consider the example
\begin{align}
\hat{C}_{4z} | 0\,0\,\uparrow_2\,\uparrow_1\rangle &= \hat{C}_{4z} c_{2,\uparrow}^{\dagger} c_{1,\uparrow}^{\dagger} | \text{vac} \rangle \nonumber \\
                                    &= c^{\dagger}_{C_{4z}^{-1}(2), \uparrow} c_{C_{4z}^{-1}(1), \uparrow}^{\dagger} | \text{vac} \rangle
                                    = c_{1,\uparrow}^{\dagger} c_{3,\uparrow}^{\dagger} | \text{vac} \rangle \nonumber \\
                                                   &= - | 0 \, \uparrow_3 \, 0\,\uparrow_1 \rangle,
\end{align}
which shows how the matrix elements of $\hat{P}_{R}$ can be constructed. 
In the particle number block $(N_{\uparrow}, N_{\downarrow})$, 
the operators $\hat{P}_R$ are permutation matrices of size 
$\text{dim}(N_{\uparrow}, N_{\downarrow}) \times \text{dim}(N_{\uparrow}, N_{\downarrow})$,
where $\text{dim}(N_{\uparrow}, N_{\downarrow}) = \begin{pmatrix} N_s \\ N_{\uparrow} \end{pmatrix} \cdot \begin{pmatrix} N_s \\ N_{\downarrow} \end{pmatrix}$,
with an additional sign structure coming from the fermionic exchanges.
Since the spatial symmetry operations do not affect the spin states, it is convenient to  write 
$\hat{P}_R$ as the tensor product $\hat{P}_R = \hat{P}_{R,\uparrow} \otimes \hat{P}_{R,\downarrow}$
with $\hat{P}_{R,\sigma}$ acting only on creation operators of spin $\sigma$.
Having obtained a matrix representation of the symmetry operators $\hat{P}_R$
on the particle number sector $(N_{\uparrow}, N_{\downarrow})$,
we can decompose this subspace of Hilbert space further into the 
irreducible invariant subspaces of $D_{4}$ via the projection operator 
technique \cite{TinkhamGroupTheory} (see also \cite{Kuns2011} for a detailed discussion).
In the decomposition
of a reducible representation 
the $n$-th irreducible representation occurs $a_n$ times, given by \cite{TinkhamGroupTheory}
\begin{equation}
 a_n = \frac{1}{h} \sum_{R} \chi^{(n)}(R)^{\star} \chi(R),
 \label{eq:n_occur_irrep}
\end{equation}
where $\chi^{(n)}(R)$ is the character of the group element $R$ 
in the $n$-th irreducible representation and $\chi(R) \equiv \text{Tr}\left(P_R\right) = \sum_{i} \left[P_R\right]_{ii}$
is the character of $R$ in the reducible matrix representation. 
Applying the formula \eqref{eq:n_occur_irrep} to each particle number sector $(N_{\uparrow}, N_{\downarrow})$
of a square plaquette we obtain the group structure 
presented in Tab.~\ref{tab:group_structure}. 

\begin{table*}
\centering
\begin{ruledtabular}
\begin{tabular}{@{}lcc@{}}
 $N_{\uparrow}, N_{\downarrow}$ & $\text{Dimension}= \begin{pmatrix} N_s \\ N_{\uparrow} \end{pmatrix} \cdot \begin{pmatrix} N_s \\ N_{\downarrow} \end{pmatrix}$ & Irreducible representations   \\ 
 \hline
 \hline
 $0,0$ & $1$ &   $A_1$    \\
 $1,0$ & $4$ &   $A_1 \oplus B_2 \oplus E$   \\
 $2,0$ & $6$ &   $A_2 \oplus B_2 \oplus 2E $  \\
 $1,1$ & $16$ &  $3 A_1 \oplus A_2 \oplus B_1 \oplus 3 B_2 \oplus 4E $  \\
 $3,0$ & $4$ &   $A_2 \oplus B_1 \oplus E $  \\
 $2,1$ & $24$ &  $3 A_1 \oplus 3 A_2 \oplus 3 B_1 \oplus 3B_2 \oplus 6E $  \\
 $4,0$ & $1$ &   $B_1 $ \\
 $3,1$ & $16$ &  $A_1 \oplus 3A_2 \oplus 3B_1 \oplus B_2 \oplus 4E $  \\
 $2,2$ & $36$ &  $6A_1 \oplus 4A_2 \oplus 6B_1 \oplus 4B_2 \oplus 8 E $ \\ 
\end{tabular}
\end{ruledtabular}
\caption[Group structure of the Hilbert space for a single square plaquette of the Hubbard model]
{Group structure of the Hilbert space for a single square plaquette of the Hubbard model.
Shown is the reduction of the subspaces of fixed particle number $(N_{\uparrow}, N_{\downarrow})$
into irreducible invariant subspaces of the symmetry group $D_4$. The table is symmetric under exchange 
of $N_{\uparrow}$ and $N_{\downarrow}$. Particle number sectors $(N_{\uparrow}, N_{\downarrow})$
above half filling have the same group structure as their particle-hole symmetric counterparts
$(N_s - N_{\uparrow}, N_s - N_{\downarrow})$ with $N_s=4$.}
\label{tab:group_structure}
\end{table*}

We wish to decompose the particle number sector $(N_{\uparrow}, N_{\downarrow})$
into blocks of states such that the application of a lattice symmetry operation
to a state mixes only states within the same block. 

Let $| \phi_{i\lambda}^{(n)} \rangle$ denote a normalized basis state that transforms according 
to the $\lambda$-th copy of the $i$-th row in the $n$-th irreducible representation. 
Then each occupation number state $|\alpha\rangle$ can be expanded as
\begin{equation}
 | \alpha \rangle = \sum_{n=1}^{c} \sum_{i=1}^{l_n} \sum_{\lambda=1}^{a_n} b_{i \lambda}^{(n)} | \phi_{i\lambda}^{(n)} \rangle,
\end{equation}
where $c$ is the number of irreducible representations, which is equal to the number of conjugacy classes \cite{TinkhamGroupTheory}
(here, for $D_4$, c=5), $l_n$ is the dimension of the $n$-th irreducible representation, and 
$\lambda$ labels the $a_n$ different copies of the $n$-th irreducible representation. 

The symmetry transfer operator is defined as \cite{TinkhamGroupTheory}
\begin{equation}
 \mathcal{P}_{ij}^{(n)} = \frac{l_n}{h} \sum_{R} \Gamma^{(n)}(R)_{ij}^{\star} P_R,
 \label{eq:symmetry_transfer_operator}
\end{equation}
where $\Gamma^{(n)}(R)$ is the matrix representation of the group element $R$
in the $n$-th irreducible representation and the sum runs over all group elements. 
$\mathcal{P}_{ii}^{(n)}$ acts as a projector onto the $i$-th row of the $n$-th 
irreducible representation, while $\mathcal{P}_{ij}^{(n)}$ transfers the $i$-th 
row into the $j$-th row according to \cite{TinkhamGroupTheory}
\begin{equation}
 \mathcal{P}_{ij}^{(n)} | \phi_{k \lambda}^{(m)} \rangle =
 \begin{cases}
 | \phi_{j \lambda}^{(n)} \rangle & \text{if } i=k \text{ and } n=m, \\  
 0 & \text{else,}
 \end{cases}
\end{equation}
and 
\begin{equation}
 \mathcal{P}_{ii}^{(n)} | \phi_{i\lambda}^{(n)} \rangle = | \phi_{i \lambda}^{(n)} \rangle.
\end{equation}
Note that if there are several copies $\lambda$ of the same irreducible representation $n$, then
the projection operator $\mathcal{P}_{ii}^{(n)}$  applied to a basis  state $|\alpha \rangle$
will return a basis state for only a single copy $\lambda(\alpha)$:
\begin{equation}
 \mathcal{P}_{ii}^{(n)} | \alpha \rangle \sim |\phi_{i \lambda(\alpha)}^{(n)} \rangle.
\end{equation}
By letting  $\mathcal{P}_{ii}^{(n)}$ act onto each state $| \alpha \rangle$ 
of the particle number sector $(N_{\uparrow}, N_{\downarrow})$ and collecting all non-zero states 
that are linearly independent, all copies of the $n$-th 
irreducible representation are generated. 
In this projection method, it may happen that the same basis state (up to a global phase) is generated multiple times. 
Thus, the basis vectors of all irreducible representations can be constructed 
and combined into a unitary matrix $S_{\alpha, (n,i,\lambda)} = \langle \alpha | \phi_{i \lambda}^{(n)} \rangle$
which transforms the reduced density matrix from the occupation number basis to 
the representation basis according to Eq.~\ref{eq:occ_to_rep} of the main text.


\section{Spin inversion symmetry}
\label{app:spin_inversion_symm}

For a finer symmetry labelling it is useful to implement the spin inversion symmetry 
$\mathcal{S} = \{\mathcal{E}, \sigma_h \}$ where $\sigma_h = \prod_{i \in A}\sigma_i^{x}$
flips all spins on subsystem $A$.
All symmetry operations of the lattice symmetry group $D_4$ commute with the spin inversion operation
since they act onto different degrees of freedom (site indices of creation operators
on the one hand and spin indices on the other hand). 
Therefore, we can form the direct-product group $D_{4h} = D_4 \times \mathcal{S}$
with 16 group elements, the original 8 from $D_4$, each multiplied
by the identity or by spin inversion, and organize the states into the irreducible invariant subspaces of $D_{4h}$.
In order to apply the projection operator method for generating the irreducible basis 
states, one needs to know the irreducible representation matrices of $D_{4h}$
(see Eq.\eqref{eq:symmetry_transfer_operator}). It can be shown \cite{TinkhamGroupTheory}
that the direct product of two irreducible representations forms an irreducible representation
of the direct product group.

If the Hilbert space is first decomposed into subspaces of fixed particle number 
$H = \prod_{\oplus \\N_{\uparrow},N_{\downarrow}=0}^{N_s} H_{(N_{\uparrow}, N_{\downarrow})}$,
then spin inversion symmetry $\mathcal{S}$ can only be used for further block diagonalization inside subspaces
with equal spin populations since it is obviously not possible to construct eigenstates of $\sigma_h$
that lie only in $H_{(N_{\uparrow}, N_{\downarrow})}$ whenever $N_{\uparrow} \ne N_{\downarrow}$.
Including spin inversion leads to the finer group 
structure of the subspaces with $N_{\uparrow}=N_{\downarrow}=1$
and $N_{\uparrow}=N_{\downarrow}=2$ 
shown in Tab.~\ref{tab:group_structure_invsym}
where the additional label $g$ ($u$) indicates 
whether the basis function is even (odd) under spin inversion.

\begin{table*}
\centering
\begin{ruledtabular}
\begin{tabular}{@{}lcc@{}}
$N_{\uparrow}, N_{\downarrow}$  & Irreducible representations   \\ 
$1,1$ & $3A_{1,g} \oplus B_{1,g} \oplus 2 B_{2,g} \oplus 2 E_{g} \oplus A_{2,u} \oplus B_{2,u} \oplus 2E_{u}$  \\
$2,2$ & $5 A_{1,g} \oplus A_{2,g} \oplus 4 B_{1,g} \oplus 3 B_{2,g} 
                                     \oplus 4 E_{g} \oplus A_{1,u} \oplus 3 A_{2,u} \oplus 2 B_{1,u} \oplus B_{2,u} \oplus 4 E_{u}$  \\
\end{tabular}
\end{ruledtabular}
\caption[Group structure of the spin-balanced subspaces of the 
single-plaquette Hubbard model.]
{Group structure of the spin-balanced subspaces of the 
single-plaquette Hubbard model. The decomposition of the subspaces 
is done with respect to the irreducible invariant subspaces of $D_{4h}$.
The subscript $g$ ($u$) denotes a wave function that is even 
(odd) under flipping all spins on the plaquette.}
\label{tab:group_structure_invsym}
\end{table*}

\section{Pseudocode for computing $\langle \beta | \rho_A | \alpha \rangle_{\{ \vecbf{s} \}}$}
\label{app:pseudocode}
The following code listing describes how to compute a matrix element 
of the reduced density matrix $\rho_A$ for a free fermion system in the 
external potential of auxiliary fields $\langle \vecbf{s} \rangle$ given the single particle
Green's function on subsystem $A$ as input. The main task consists in collecting 
the appropriate row and column indices for the submatrices, which appear in the 
determinant formula of Eq.~\eqref{eq:Wicks_theorem_general}.
The symbols $\mathcal{I}_{+}$ and $\mathcal{I}_{-}$ and the meaning of the 
abbreviations "occ`` and ''ua`` are defined in the main text. 
The phase factors in line 23 are those from Eqs.~\eqref{eq:minus1_p} and \eqref{eq:minus1_p_prime}
in the main text. Note that the computation of a many-body reduced density matrix 
from a single-particle Green's function obviously constitutes a blow-up of redundant 
information. 
\begin{figure}
\begin{algorithm}[H]
 \caption{Reduced density matrix}
\label{alg:rhoA_alpha_beta}
{\bf Result:} 
 Matrix element $\langle \beta | \rho_A | \alpha \rangle_{\{ \vecbf{s} \}}$ between the occupation number states $|\alpha \rangle$
              and $|\beta \rangle$.
\par \noindent             
{\bf Input:}
\begin{itemize}
 \item Occupation states $|\alpha \rangle = |\alpha_{\uparrow} \rangle \otimes | \alpha_{\downarrow} \rangle$
       and $|\beta \rangle = |\beta_{\uparrow} \rangle \otimes | \beta_{\downarrow} \rangle$
       bitcoded as integers $[\alpha_{\uparrow}],[\alpha_{\downarrow}],[\beta_{\uparrow}],[\beta_{\downarrow}]$  
 \item Single-particle Green's function $G^{(0)}(1:N_{\text{sites},A},1:N_{\text{sites},A}; \sigma=\uparrow,\downarrow)$ 
       for Hubbard-Stratonovich configuration $\{\vecbf{s}\}$, restricted to subsystem $A$.
\end{itemize}

\begin{algorithmic}[1]

\For{$\sigma = \uparrow, \downarrow$}

\State $[t_1] = \text{XOR}([\alpha_{\sigma}], [\beta_{\sigma}])$ 
\State $[t_2] = \text{NOT}([t_1])$
\State $[t_{-}] = \text{AND}([t_1], [\alpha_{\sigma}])$
\State $\quad \mathcal{I}_{-} = \text{bitonesToSitelist}([t_{-}])$
\State $[t_{+}] = \text{NOT}(\text{AND}([t_1], [\beta_{\sigma}]))$
\State $\quad \mathcal{I}_{+} = \text{bitonesToSitelist}([t_{+}])$
\State $[t_3] = \text{AND}([\alpha_{\sigma}],[t_2])$
\State $\quad \mathcal{I}_{\text{occ, ua}} = \text{bitonesToSitelist}([t_3])$
\State $N_{\text{occ}} = |\mathcal{I}_{\text{occ, ua}}|$

\State $r_{\sigma} = 0$
\For{$b = 0:N_{\text{occ}}$}
\Comment{``branches'' for occ. sites}

\State $\mathcal{I}_{\text{occ, ua}}^{\text{branch}} = \{ i_k \in \mathcal{I}_{\text{occ, ua}} \,|\, k\text{-th bit in } [b] \text{ is set.}  \}$
\State $N_{\text{BitonesBranch}} = |\mathcal{I}_{\text{occ, ua}}^{\text{branch}}|$
\State $\mathcal{I}_{\text{proj}} = \mathcal{I}_{\text{occ, ua}}^{\text{branch}} 
       \cup \mathcal{I}_{\text{unocc, ua}}$
\State $\mathcal{R} = \mathcal{I}_{\text{proj}} \cup \mathcal{I}_{-}$\Comment{List of sites for row indices}
\State $\mathcal{C} = \mathcal{I}_{\text{proj}} \cup \mathcal{I}_{+}$\Comment{List of sites for column indices}
\State $k=|\mathcal{C}| (=|\mathcal{R}|)$
\For{$i=1:k$}
\For{$j=1:k$}
\State $G_k^{(0)}(i,j) =  G^{(0)}(\mathcal{R}(i), \mathcal{C}(j); \sigma)$
\EndFor
\EndFor
\State $r_{\sigma} = r_{\sigma} +  (-1)^{N_{\text{BitonesBranch}}} \det\left( G_k^{(0)} \right)$

\EndFor
\EndFor

\State $\langle \beta | \rho_A | \alpha \rangle_{\{\vecbf{s}\}} = (-1)^{p_{\uparrow}} (-1)^{p_{\uparrow}^{\prime}} (-1)^{p_{\downarrow}} (-1)^{p_{\downarrow}^{\prime}} r_{\uparrow} \cdot r_{\downarrow} $

\end{algorithmic}
\par \noindent 
{\bf Notation:} $[x]$ means that the integer $x$ is to be replaced by its binary bit string,
where each bit indicates the occupation of a lattice site. 
$|\mathcal{I}|$ denotes the number of elements in the list $\mathcal{I}$ and 
$\text{bitonesTopSitelist}([x])$ is a routine that returns a list of lattice sites corresponding 
to the positions in the bitstring $[x]$ where the bit is set. \\
\end{algorithm}
\end{figure}

\pagebreak 


\begin{thebibliography}{52}%
\makeatletter
\providecommand \@ifxundefined [1]{%
 \@ifx{#1\undefined}
}%
\providecommand \@ifnum [1]{%
 \ifnum #1\expandafter \@firstoftwo
 \else \expandafter \@secondoftwo
 \fi
}%
\providecommand \@ifx [1]{%
 \ifx #1\expandafter \@firstoftwo
 \else \expandafter \@secondoftwo
 \fi
}%
\providecommand \natexlab [1]{#1}%
\providecommand \enquote  [1]{``#1''}%
\providecommand \bibnamefont  [1]{#1}%
\providecommand \bibfnamefont [1]{#1}%
\providecommand \citenamefont [1]{#1}%
\providecommand \href@noop [0]{\@secondoftwo}%
\providecommand \href [0]{\begingroup \@sanitize@url \@href}%
\providecommand \@href[1]{\@@startlink{#1}\@@href}%
\providecommand \@@href[1]{\endgroup#1\@@endlink}%
\providecommand \@sanitize@url [0]{\catcode `\\12\catcode `\$12\catcode
  `\&12\catcode `\#12\catcode `\^12\catcode `\_12\catcode `\%12\relax}%
\providecommand \@@startlink[1]{}%
\providecommand \@@endlink[0]{}%
\providecommand \url  [0]{\begingroup\@sanitize@url \@url }%
\providecommand \@url [1]{\endgroup\@href {#1}{\urlprefix }}%
\providecommand \urlprefix  [0]{URL }%
\providecommand \Eprint [0]{\href }%
\providecommand \doibase [0]{http://dx.doi.org/}%
\providecommand \selectlanguage [0]{\@gobble}%
\providecommand \bibinfo  [0]{\@secondoftwo}%
\providecommand \bibfield  [0]{\@secondoftwo}%
\providecommand \translation [1]{[#1]}%
\providecommand \BibitemOpen [0]{}%
\providecommand \bibitemStop [0]{}%
\providecommand \bibitemNoStop [0]{.\EOS\space}%
\providecommand \EOS [0]{\spacefactor3000\relax}%
\providecommand \BibitemShut  [1]{\csname bibitem#1\endcsname}%
\let\auto@bib@innerbib\@empty
\bibitem [{\citenamefont {H{\"a}ffner}\ \emph {et~al.}(2005)\citenamefont
  {H{\"a}ffner}, \citenamefont {H{\"a}nsel}, \citenamefont {Roos},
  \citenamefont {Benhelm}, \citenamefont {Chwalla}, \citenamefont {K{\"o}rber},
  \citenamefont {Rapol}, \citenamefont {Riebe}, \citenamefont {Schmidt},
  \citenamefont {Becher} \emph {et~al.}}]{Haeffner2005}%
  \BibitemOpen
  \bibfield  {author} {\bibinfo {author} {\bibfnamefont {H.}~\bibnamefont
  {H{\"a}ffner}}, \bibinfo {author} {\bibfnamefont {W.}~\bibnamefont
  {H{\"a}nsel}}, \bibinfo {author} {\bibfnamefont {C.}~\bibnamefont {Roos}},
  \bibinfo {author} {\bibfnamefont {J.}~\bibnamefont {Benhelm}}, \bibinfo
  {author} {\bibfnamefont {M.}~\bibnamefont {Chwalla}}, \bibinfo {author}
  {\bibfnamefont {T.}~\bibnamefont {K{\"o}rber}}, \bibinfo {author}
  {\bibfnamefont {U.}~\bibnamefont {Rapol}}, \bibinfo {author} {\bibfnamefont
  {M.}~\bibnamefont {Riebe}}, \bibinfo {author} {\bibfnamefont
  {P.}~\bibnamefont {Schmidt}}, \bibinfo {author} {\bibfnamefont
  {C.}~\bibnamefont {Becher}},  \emph {et~al.},\ }\href@noop {} {\bibfield
  {journal} {\bibinfo  {journal} {Nature}\ }\textbf {\bibinfo {volume} {438}},\
  \bibinfo {pages} {643} (\bibinfo {year} {2005})}\BibitemShut {NoStop}%
\bibitem [{\citenamefont {Xin}\ \emph {et~al.}(2017)\citenamefont {Xin},
  \citenamefont {Lu}, \citenamefont {Klassen}, \citenamefont {Yu},
  \citenamefont {Ji}, \citenamefont {Chen}, \citenamefont {Ma}, \citenamefont
  {Long}, \citenamefont {Zeng},\ and\ \citenamefont {Laflamme}}]{TaoXin2017}%
  \BibitemOpen
  \bibfield  {author} {\bibinfo {author} {\bibfnamefont {T.}~\bibnamefont
  {Xin}}, \bibinfo {author} {\bibfnamefont {D.}~\bibnamefont {Lu}}, \bibinfo
  {author} {\bibfnamefont {J.}~\bibnamefont {Klassen}}, \bibinfo {author}
  {\bibfnamefont {N.}~\bibnamefont {Yu}}, \bibinfo {author} {\bibfnamefont
  {Z.}~\bibnamefont {Ji}}, \bibinfo {author} {\bibfnamefont {J.}~\bibnamefont
  {Chen}}, \bibinfo {author} {\bibfnamefont {X.}~\bibnamefont {Ma}}, \bibinfo
  {author} {\bibfnamefont {G.}~\bibnamefont {Long}}, \bibinfo {author}
  {\bibfnamefont {B.}~\bibnamefont {Zeng}}, \ and\ \bibinfo {author}
  {\bibfnamefont {R.}~\bibnamefont {Laflamme}},\ }\href {\doibase
  10.1103/PhysRevLett.118.020401} {\bibfield  {journal} {\bibinfo  {journal}
  {Phys. Rev. Lett.}\ }\textbf {\bibinfo {volume} {118}},\ \bibinfo {pages}
  {020401} (\bibinfo {year} {2017})}\BibitemShut {NoStop}%
\bibitem [{\citenamefont {Baur}\ \emph {et~al.}(2012)\citenamefont {Baur},
  \citenamefont {Fedorov}, \citenamefont {Steffen}, \citenamefont {Filipp},
  \citenamefont {da~Silva},\ and\ \citenamefont {Wallraff}}]{Baur2012}%
  \BibitemOpen
  \bibfield  {author} {\bibinfo {author} {\bibfnamefont {M.}~\bibnamefont
  {Baur}}, \bibinfo {author} {\bibfnamefont {A.}~\bibnamefont {Fedorov}},
  \bibinfo {author} {\bibfnamefont {L.}~\bibnamefont {Steffen}}, \bibinfo
  {author} {\bibfnamefont {S.}~\bibnamefont {Filipp}}, \bibinfo {author}
  {\bibfnamefont {M.~P.}\ \bibnamefont {da~Silva}}, \ and\ \bibinfo {author}
  {\bibfnamefont {A.}~\bibnamefont {Wallraff}},\ }\href {\doibase
  10.1103/PhysRevLett.108.040502} {\bibfield  {journal} {\bibinfo  {journal}
  {Phys. Rev. Lett.}\ }\textbf {\bibinfo {volume} {108}},\ \bibinfo {pages}
  {040502} (\bibinfo {year} {2012})}\BibitemShut {NoStop}%
\bibitem [{\citenamefont {Schwemmer}\ \emph {et~al.}(2014)\citenamefont
  {Schwemmer}, \citenamefont {T\'oth}, \citenamefont {Niggebaum}, \citenamefont
  {Moroder}, \citenamefont {Gross}, \citenamefont {G\"uhne},\ and\
  \citenamefont {Weinfurter}}]{Schwemmer2014}%
  \BibitemOpen
  \bibfield  {author} {\bibinfo {author} {\bibfnamefont {C.}~\bibnamefont
  {Schwemmer}}, \bibinfo {author} {\bibfnamefont {G.}~\bibnamefont {T\'oth}},
  \bibinfo {author} {\bibfnamefont {A.}~\bibnamefont {Niggebaum}}, \bibinfo
  {author} {\bibfnamefont {T.}~\bibnamefont {Moroder}}, \bibinfo {author}
  {\bibfnamefont {D.}~\bibnamefont {Gross}}, \bibinfo {author} {\bibfnamefont
  {O.}~\bibnamefont {G\"uhne}}, \ and\ \bibinfo {author} {\bibfnamefont
  {H.}~\bibnamefont {Weinfurter}},\ }\href {\doibase
  10.1103/PhysRevLett.113.040503} {\bibfield  {journal} {\bibinfo  {journal}
  {Phys. Rev. Lett.}\ }\textbf {\bibinfo {volume} {113}},\ \bibinfo {pages}
  {040503} (\bibinfo {year} {2014})}\BibitemShut {NoStop}%
\bibitem [{\citenamefont {Gao}\ \emph {et~al.}(2018)\citenamefont {Gao},
  \citenamefont {Qiao}, \citenamefont {Jiao}, \citenamefont {Ma}, \citenamefont
  {Hu}, \citenamefont {Ren}, \citenamefont {Yang}, \citenamefont {Tang},
  \citenamefont {Yung},\ and\ \citenamefont {Jin}}]{JunGao2018}%
  \BibitemOpen
  \bibfield  {author} {\bibinfo {author} {\bibfnamefont {J.}~\bibnamefont
  {Gao}}, \bibinfo {author} {\bibfnamefont {L.-F.}\ \bibnamefont {Qiao}},
  \bibinfo {author} {\bibfnamefont {Z.-Q.}\ \bibnamefont {Jiao}}, \bibinfo
  {author} {\bibfnamefont {Y.-C.}\ \bibnamefont {Ma}}, \bibinfo {author}
  {\bibfnamefont {C.-Q.}\ \bibnamefont {Hu}}, \bibinfo {author} {\bibfnamefont
  {R.-J.}\ \bibnamefont {Ren}}, \bibinfo {author} {\bibfnamefont {A.-L.}\
  \bibnamefont {Yang}}, \bibinfo {author} {\bibfnamefont {H.}~\bibnamefont
  {Tang}}, \bibinfo {author} {\bibfnamefont {M.-H.}\ \bibnamefont {Yung}}, \
  and\ \bibinfo {author} {\bibfnamefont {X.-M.}\ \bibnamefont {Jin}},\ }\href
  {\doibase 10.1103/PhysRevLett.120.240501} {\bibfield  {journal} {\bibinfo
  {journal} {Phys. Rev. Lett.}\ }\textbf {\bibinfo {volume} {120}},\ \bibinfo
  {pages} {240501} (\bibinfo {year} {2018})}\BibitemShut {NoStop}%
\bibitem [{\citenamefont {Gross}\ \emph {et~al.}(2010)\citenamefont {Gross},
  \citenamefont {Liu}, \citenamefont {Flammia}, \citenamefont {Becker},\ and\
  \citenamefont {Eisert}}]{Gross2010}%
  \BibitemOpen
  \bibfield  {author} {\bibinfo {author} {\bibfnamefont {D.}~\bibnamefont
  {Gross}}, \bibinfo {author} {\bibfnamefont {Y.-K.}\ \bibnamefont {Liu}},
  \bibinfo {author} {\bibfnamefont {S.~T.}\ \bibnamefont {Flammia}}, \bibinfo
  {author} {\bibfnamefont {S.}~\bibnamefont {Becker}}, \ and\ \bibinfo {author}
  {\bibfnamefont {J.}~\bibnamefont {Eisert}},\ }\href {\doibase
  10.1103/PhysRevLett.105.150401} {\bibfield  {journal} {\bibinfo  {journal}
  {Phys. Rev. Lett.}\ }\textbf {\bibinfo {volume} {105}},\ \bibinfo {pages}
  {150401} (\bibinfo {year} {2010})}\BibitemShut {NoStop}%
\bibitem [{\citenamefont {Cramer}\ \emph {et~al.}(2010)\citenamefont {Cramer},
  \citenamefont {Plenio}, \citenamefont {Flammia}, \citenamefont {Somma},
  \citenamefont {Gross}, \citenamefont {Bartlett}, \citenamefont
  {Landon-Cardinal}, \citenamefont {Poulin},\ and\ \citenamefont
  {Liu}}]{Cramer2010}%
  \BibitemOpen
  \bibfield  {author} {\bibinfo {author} {\bibfnamefont {M.}~\bibnamefont
  {Cramer}}, \bibinfo {author} {\bibfnamefont {M.~B.}\ \bibnamefont {Plenio}},
  \bibinfo {author} {\bibfnamefont {S.~T.}\ \bibnamefont {Flammia}}, \bibinfo
  {author} {\bibfnamefont {R.}~\bibnamefont {Somma}}, \bibinfo {author}
  {\bibfnamefont {D.}~\bibnamefont {Gross}}, \bibinfo {author} {\bibfnamefont
  {S.~D.}\ \bibnamefont {Bartlett}}, \bibinfo {author} {\bibfnamefont
  {O.}~\bibnamefont {Landon-Cardinal}}, \bibinfo {author} {\bibfnamefont
  {D.}~\bibnamefont {Poulin}}, \ and\ \bibinfo {author} {\bibfnamefont {Y.-K.}\
  \bibnamefont {Liu}},\ }\href@noop {} {\bibfield  {journal} {\bibinfo
  {journal} {Nature communications}\ }\textbf {\bibinfo {volume} {1}},\
  \bibinfo {pages} {149} (\bibinfo {year} {2010})}\BibitemShut {NoStop}%
\bibitem [{\citenamefont {Riofr{\'\i}o}\ \emph {et~al.}(2017)\citenamefont
  {Riofr{\'\i}o}, \citenamefont {Gross}, \citenamefont {Flammia}, \citenamefont
  {Monz}, \citenamefont {Nigg}, \citenamefont {Blatt},\ and\ \citenamefont
  {Eisert}}]{Riofrio2017}%
  \BibitemOpen
  \bibfield  {author} {\bibinfo {author} {\bibfnamefont {C.}~\bibnamefont
  {Riofr{\'\i}o}}, \bibinfo {author} {\bibfnamefont {D.}~\bibnamefont {Gross}},
  \bibinfo {author} {\bibfnamefont {S.}~\bibnamefont {Flammia}}, \bibinfo
  {author} {\bibfnamefont {T.}~\bibnamefont {Monz}}, \bibinfo {author}
  {\bibfnamefont {D.}~\bibnamefont {Nigg}}, \bibinfo {author} {\bibfnamefont
  {R.}~\bibnamefont {Blatt}}, \ and\ \bibinfo {author} {\bibfnamefont
  {J.}~\bibnamefont {Eisert}},\ }\href@noop {} {\bibfield  {journal} {\bibinfo
  {journal} {Nature communications}\ }\textbf {\bibinfo {volume} {8}},\
  \bibinfo {pages} {15305} (\bibinfo {year} {2017})}\BibitemShut {NoStop}%
\bibitem [{\citenamefont {Steffens}\ \emph {et~al.}(2015)\citenamefont
  {Steffens}, \citenamefont {Friesdorf}, \citenamefont {Langen}, \citenamefont
  {Rauer}, \citenamefont {Schweigler}, \citenamefont {H{\"u}bener},
  \citenamefont {Schmiedmayer}, \citenamefont {Riofr{\'\i}o},\ and\
  \citenamefont {Eisert}}]{Steffens2015}%
  \BibitemOpen
  \bibfield  {author} {\bibinfo {author} {\bibfnamefont {A.}~\bibnamefont
  {Steffens}}, \bibinfo {author} {\bibfnamefont {M.}~\bibnamefont {Friesdorf}},
  \bibinfo {author} {\bibfnamefont {T.}~\bibnamefont {Langen}}, \bibinfo
  {author} {\bibfnamefont {B.}~\bibnamefont {Rauer}}, \bibinfo {author}
  {\bibfnamefont {T.}~\bibnamefont {Schweigler}}, \bibinfo {author}
  {\bibfnamefont {R.}~\bibnamefont {H{\"u}bener}}, \bibinfo {author}
  {\bibfnamefont {J.}~\bibnamefont {Schmiedmayer}}, \bibinfo {author}
  {\bibfnamefont {C.}~\bibnamefont {Riofr{\'\i}o}}, \ and\ \bibinfo {author}
  {\bibfnamefont {J.}~\bibnamefont {Eisert}},\ }\href@noop {} {\bibfield
  {journal} {\bibinfo  {journal} {Nature communications}\ }\textbf {\bibinfo
  {volume} {6}},\ \bibinfo {pages} {7663} (\bibinfo {year} {2015})}\BibitemShut
  {NoStop}%
\bibitem [{\citenamefont {{Mazurenko}}\ \emph {et~al.}(2017)\citenamefont
  {{Mazurenko}}, \citenamefont {{Chiu}}, \citenamefont {{Ji}}, \citenamefont
  {{Parsons}}, \citenamefont {{Kan{\'a}sz-Nagy}}, \citenamefont {{Schmidt}},
  \citenamefont {{Grusdt}}, \citenamefont {{Demler}}, \citenamefont {{Greif}},\
  and\ \citenamefont {{Greiner}}}]{Mazurenko2017}%
  \BibitemOpen
  \bibfield  {author} {\bibinfo {author} {\bibfnamefont {A.}~\bibnamefont
  {{Mazurenko}}}, \bibinfo {author} {\bibfnamefont {C.~S.}\ \bibnamefont
  {{Chiu}}}, \bibinfo {author} {\bibfnamefont {G.}~\bibnamefont {{Ji}}},
  \bibinfo {author} {\bibfnamefont {M.~F.}\ \bibnamefont {{Parsons}}}, \bibinfo
  {author} {\bibfnamefont {M.}~\bibnamefont {{Kan{\'a}sz-Nagy}}}, \bibinfo
  {author} {\bibfnamefont {R.}~\bibnamefont {{Schmidt}}}, \bibinfo {author}
  {\bibfnamefont {F.}~\bibnamefont {{Grusdt}}}, \bibinfo {author}
  {\bibfnamefont {E.}~\bibnamefont {{Demler}}}, \bibinfo {author}
  {\bibfnamefont {D.}~\bibnamefont {{Greif}}}, \ and\ \bibinfo {author}
  {\bibfnamefont {M.}~\bibnamefont {{Greiner}}},\ }\href@noop {} {\bibfield
  {journal} {\bibinfo  {journal} {Nature}\ }\textbf {\bibinfo {volume} {545}},\
  \bibinfo {pages} {462} (\bibinfo {year} {2017})}\BibitemShut {NoStop}%
\bibitem [{\citenamefont {Cheuk}\ \emph {et~al.}(2016)\citenamefont {Cheuk},
  \citenamefont {Nichols}, \citenamefont {Lawrence}, \citenamefont {Okan},
  \citenamefont {Zhang}, \citenamefont {Khatami}, \citenamefont {Trivedi},
  \citenamefont {Paiva}, \citenamefont {Rigol},\ and\ \citenamefont
  {Zwierlein}}]{Cheuk2016}%
  \BibitemOpen
  \bibfield  {author} {\bibinfo {author} {\bibfnamefont {L.~W.}\ \bibnamefont
  {Cheuk}}, \bibinfo {author} {\bibfnamefont {M.~A.}\ \bibnamefont {Nichols}},
  \bibinfo {author} {\bibfnamefont {K.~R.}\ \bibnamefont {Lawrence}}, \bibinfo
  {author} {\bibfnamefont {M.}~\bibnamefont {Okan}}, \bibinfo {author}
  {\bibfnamefont {H.}~\bibnamefont {Zhang}}, \bibinfo {author} {\bibfnamefont
  {E.}~\bibnamefont {Khatami}}, \bibinfo {author} {\bibfnamefont
  {N.}~\bibnamefont {Trivedi}}, \bibinfo {author} {\bibfnamefont
  {T.}~\bibnamefont {Paiva}}, \bibinfo {author} {\bibfnamefont
  {M.}~\bibnamefont {Rigol}}, \ and\ \bibinfo {author} {\bibfnamefont {M.~W.}\
  \bibnamefont {Zwierlein}},\ }\href {\doibase 10.1126/science.aag3349}
  {\bibfield  {journal} {\bibinfo  {journal} {Science}\ }\textbf {\bibinfo
  {volume} {353}},\ \bibinfo {pages} {1260} (\bibinfo {year} {2016})},\ \Eprint
  {http://arxiv.org/abs/http://science.sciencemag.org/content/353/6305/1260.full.pdf}
  {http://science.sciencemag.org/content/353/6305/1260.full.pdf} \BibitemShut
  {NoStop}%
\bibitem [{\citenamefont {{Mitra}}\ \emph {et~al.}(2017)\citenamefont
  {{Mitra}}, \citenamefont {{Brown}}, \citenamefont {{Guardado-Sanchez}},
  \citenamefont {{Kondov}}, \citenamefont {{Devakul}}, \citenamefont {{Huse}},
  \citenamefont {{Schauss}},\ and\ \citenamefont {{Bakr}}}]{Mitra2017}%
  \BibitemOpen
  \bibfield  {author} {\bibinfo {author} {\bibfnamefont {D.}~\bibnamefont
  {{Mitra}}}, \bibinfo {author} {\bibfnamefont {P.~T.}\ \bibnamefont
  {{Brown}}}, \bibinfo {author} {\bibfnamefont {E.}~\bibnamefont
  {{Guardado-Sanchez}}}, \bibinfo {author} {\bibfnamefont {S.~S.}\ \bibnamefont
  {{Kondov}}}, \bibinfo {author} {\bibfnamefont {T.}~\bibnamefont {{Devakul}}},
  \bibinfo {author} {\bibfnamefont {D.~A.}\ \bibnamefont {{Huse}}}, \bibinfo
  {author} {\bibfnamefont {P.}~\bibnamefont {{Schauss}}}, \ and\ \bibinfo
  {author} {\bibfnamefont {W.~S.}\ \bibnamefont {{Bakr}}},\ }\href@noop {}
  {\bibfield  {journal} {\bibinfo  {journal} {Nature Physics}\ }\textbf
  {\bibinfo {volume} {14}},\ \bibinfo {pages} {173} (\bibinfo {year} {2017})},\
  \Eprint {http://arxiv.org/abs/1705.02039} {arXiv:1705.02039
  [cond-mat.quant-gas]} \BibitemShut {NoStop}%
\bibitem [{\citenamefont {{Koepsell}}\ \emph {et~al.}(2018)\citenamefont
  {{Koepsell}}, \citenamefont {{Vijayan}}, \citenamefont {{Sompet}},
  \citenamefont {{Grusdt}}, \citenamefont {{Hilker}}, \citenamefont {{Demler}},
  \citenamefont {{Salomon}}, \citenamefont {{Bloch}},\ and\ \citenamefont
  {{Gross}}}]{Koepsell2018}%
  \BibitemOpen
  \bibfield  {author} {\bibinfo {author} {\bibfnamefont {J.}~\bibnamefont
  {{Koepsell}}}, \bibinfo {author} {\bibfnamefont {J.}~\bibnamefont
  {{Vijayan}}}, \bibinfo {author} {\bibfnamefont {P.}~\bibnamefont {{Sompet}}},
  \bibinfo {author} {\bibfnamefont {F.}~\bibnamefont {{Grusdt}}}, \bibinfo
  {author} {\bibfnamefont {T.~A.}\ \bibnamefont {{Hilker}}}, \bibinfo {author}
  {\bibfnamefont {E.}~\bibnamefont {{Demler}}}, \bibinfo {author}
  {\bibfnamefont {G.}~\bibnamefont {{Salomon}}}, \bibinfo {author}
  {\bibfnamefont {I.}~\bibnamefont {{Bloch}}}, \ and\ \bibinfo {author}
  {\bibfnamefont {C.}~\bibnamefont {{Gross}}},\ }\href@noop {} {\bibfield
  {journal} {\bibinfo  {journal} {arXiv e-prints}\ ,\ \bibinfo {eid}
  {arXiv:1811.06907}} (\bibinfo {year} {2018})},\ \Eprint
  {http://arxiv.org/abs/1811.06907} {arXiv:1811.06907 [cond-mat.quant-gas]}
  \BibitemShut {NoStop}%
\bibitem [{\citenamefont {Chiu}\ \emph {et~al.}(2018)\citenamefont {Chiu},
  \citenamefont {Ji}, \citenamefont {Bohrdt}, \citenamefont {Xu}, \citenamefont
  {Knap}, \citenamefont {Demler}, \citenamefont {Grusdt}, \citenamefont
  {Greiner},\ and\ \citenamefont {Greif}}]{Chiu2018}%
  \BibitemOpen
  \bibfield  {author} {\bibinfo {author} {\bibfnamefont {C.~S.}\ \bibnamefont
  {Chiu}}, \bibinfo {author} {\bibfnamefont {G.}~\bibnamefont {Ji}}, \bibinfo
  {author} {\bibfnamefont {A.}~\bibnamefont {Bohrdt}}, \bibinfo {author}
  {\bibfnamefont {M.}~\bibnamefont {Xu}}, \bibinfo {author} {\bibfnamefont
  {M.}~\bibnamefont {Knap}}, \bibinfo {author} {\bibfnamefont {E.}~\bibnamefont
  {Demler}}, \bibinfo {author} {\bibfnamefont {F.}~\bibnamefont {Grusdt}},
  \bibinfo {author} {\bibfnamefont {M.}~\bibnamefont {Greiner}}, \ and\
  \bibinfo {author} {\bibfnamefont {D.}~\bibnamefont {Greif}},\ }\href@noop {}
  {\bibfield  {journal} {\bibinfo  {journal} {arXiv preprint arXiv:1810.03584}\
  } (\bibinfo {year} {2018})}\BibitemShut {NoStop}%
\bibitem [{\citenamefont {\ifmmode~\check{S}\else \v{S}\fi{}imkovic}\ \emph
  {et~al.}(2017)\citenamefont {\ifmmode~\check{S}\else \v{S}\fi{}imkovic},
  \citenamefont {Deng}, \citenamefont {Prokof'ev}, \citenamefont {Svistunov},
  \citenamefont {Tupitsyn},\ and\ \citenamefont {Kozik}}]{Siimkovic2017}%
  \BibitemOpen
  \bibfield  {author} {\bibinfo {author} {\bibfnamefont {F.}~\bibnamefont
  {\ifmmode~\check{S}\else \v{S}\fi{}imkovic}}, \bibinfo {author}
  {\bibfnamefont {Y.}~\bibnamefont {Deng}}, \bibinfo {author} {\bibfnamefont
  {N.~V.}\ \bibnamefont {Prokof'ev}}, \bibinfo {author} {\bibfnamefont {B.~V.}\
  \bibnamefont {Svistunov}}, \bibinfo {author} {\bibfnamefont {I.~S.}\
  \bibnamefont {Tupitsyn}}, \ and\ \bibinfo {author} {\bibfnamefont
  {E.}~\bibnamefont {Kozik}},\ }\href {\doibase 10.1103/PhysRevB.96.081117}
  {\bibfield  {journal} {\bibinfo  {journal} {Phys. Rev. B}\ }\textbf {\bibinfo
  {volume} {96}},\ \bibinfo {pages} {081117(R)} (\bibinfo {year}
  {2017})}\BibitemShut {NoStop}%
\bibitem [{\citenamefont {{Humeniuk}}\ and\ \citenamefont
  {{B{\"u}chler}}(2017)}]{Humeniuk2017}%
  \BibitemOpen
  \bibfield  {author} {\bibinfo {author} {\bibfnamefont {S.}~\bibnamefont
  {{Humeniuk}}}\ and\ \bibinfo {author} {\bibfnamefont {H.~P.}\ \bibnamefont
  {{B{\"u}chler}}},\ }\href {\doibase 10.1103/PhysRevLett.119.236401}
  {\bibfield  {journal} {\bibinfo  {journal} {Physical Review Letters}\
  }\textbf {\bibinfo {volume} {119}},\ \bibinfo {eid} {236401} (\bibinfo {year}
  {2017})},\ \Eprint {http://arxiv.org/abs/1706.08951} {arXiv:1706.08951
  [cond-mat.str-el]} \BibitemShut {NoStop}%
\bibitem [{\citenamefont {Parsons}\ \emph {et~al.}(2016)\citenamefont
  {Parsons}, \citenamefont {Mazurenko}, \citenamefont {Chiu}, \citenamefont
  {Ji}, \citenamefont {Greif},\ and\ \citenamefont {Greiner}}]{Parsons2016}%
  \BibitemOpen
  \bibfield  {author} {\bibinfo {author} {\bibfnamefont {M.~F.}\ \bibnamefont
  {Parsons}}, \bibinfo {author} {\bibfnamefont {A.}~\bibnamefont {Mazurenko}},
  \bibinfo {author} {\bibfnamefont {C.~S.}\ \bibnamefont {Chiu}}, \bibinfo
  {author} {\bibfnamefont {G.}~\bibnamefont {Ji}}, \bibinfo {author}
  {\bibfnamefont {D.}~\bibnamefont {Greif}}, \ and\ \bibinfo {author}
  {\bibfnamefont {M.}~\bibnamefont {Greiner}},\ }\href@noop {} {\bibfield
  {journal} {\bibinfo  {journal} {Science}\ }\textbf {\bibinfo {volume}
  {353}},\ \bibinfo {pages} {1253} (\bibinfo {year} {2016})}\BibitemShut
  {NoStop}%
\bibitem [{\citenamefont {Daley}\ \emph {et~al.}(2012)\citenamefont {Daley},
  \citenamefont {Pichler}, \citenamefont {Schachenmayer},\ and\ \citenamefont
  {Zoller}}]{Daley2012}%
  \BibitemOpen
  \bibfield  {author} {\bibinfo {author} {\bibfnamefont {A.~J.}\ \bibnamefont
  {Daley}}, \bibinfo {author} {\bibfnamefont {H.}~\bibnamefont {Pichler}},
  \bibinfo {author} {\bibfnamefont {J.}~\bibnamefont {Schachenmayer}}, \ and\
  \bibinfo {author} {\bibfnamefont {P.}~\bibnamefont {Zoller}},\ }\href
  {\doibase 10.1103/PhysRevLett.109.020505} {\bibfield  {journal} {\bibinfo
  {journal} {Phys. Rev. Lett.}\ }\textbf {\bibinfo {volume} {109}},\ \bibinfo
  {pages} {020505} (\bibinfo {year} {2012})}\BibitemShut {NoStop}%
\bibitem [{\citenamefont {Islam}\ \emph {et~al.}(2015)\citenamefont {Islam},
  \citenamefont {Ma}, \citenamefont {Preiss}, \citenamefont {Tai},
  \citenamefont {Lukin}, \citenamefont {Rispoli},\ and\ \citenamefont
  {Greiner}}]{Islam2015}%
  \BibitemOpen
  \bibfield  {author} {\bibinfo {author} {\bibfnamefont {R.}~\bibnamefont
  {Islam}}, \bibinfo {author} {\bibfnamefont {R.}~\bibnamefont {Ma}}, \bibinfo
  {author} {\bibfnamefont {P.~M.}\ \bibnamefont {Preiss}}, \bibinfo {author}
  {\bibfnamefont {M.~E.}\ \bibnamefont {Tai}}, \bibinfo {author} {\bibfnamefont
  {A.}~\bibnamefont {Lukin}}, \bibinfo {author} {\bibfnamefont
  {M.}~\bibnamefont {Rispoli}}, \ and\ \bibinfo {author} {\bibfnamefont
  {M.}~\bibnamefont {Greiner}},\ }\href@noop {} {\bibfield  {journal} {\bibinfo
   {journal} {Nature}\ }\textbf {\bibinfo {volume} {528}},\ \bibinfo {pages}
  {77} (\bibinfo {year} {2015})}\BibitemShut {NoStop}%
\bibitem [{\citenamefont {Pichler}\ \emph {et~al.}(2013)\citenamefont
  {Pichler}, \citenamefont {Bonnes}, \citenamefont {Daley}, \citenamefont
  {Läuchli},\ and\ \citenamefont {Zoller}}]{Pichler2013}%
  \BibitemOpen
  \bibfield  {author} {\bibinfo {author} {\bibfnamefont {H.}~\bibnamefont
  {Pichler}}, \bibinfo {author} {\bibfnamefont {L.}~\bibnamefont {Bonnes}},
  \bibinfo {author} {\bibfnamefont {A.~J.}\ \bibnamefont {Daley}}, \bibinfo
  {author} {\bibfnamefont {A.~M.}\ \bibnamefont {Läuchli}}, \ and\ \bibinfo
  {author} {\bibfnamefont {P.}~\bibnamefont {Zoller}},\ }\href {\doibase
  10.1088/1367-2630/15/6/063003} {\bibfield  {journal} {\bibinfo  {journal}
  {New Journal of Physics}\ }\textbf {\bibinfo {volume} {15}},\ \bibinfo
  {pages} {063003} (\bibinfo {year} {2013})}\BibitemShut {NoStop}%
\bibitem [{\citenamefont {Pichler}\ \emph {et~al.}(2016)\citenamefont
  {Pichler}, \citenamefont {Zhu}, \citenamefont {Seif}, \citenamefont
  {Zoller},\ and\ \citenamefont {Hafezi}}]{Pichler2016}%
  \BibitemOpen
  \bibfield  {author} {\bibinfo {author} {\bibfnamefont {H.}~\bibnamefont
  {Pichler}}, \bibinfo {author} {\bibfnamefont {G.}~\bibnamefont {Zhu}},
  \bibinfo {author} {\bibfnamefont {A.}~\bibnamefont {Seif}}, \bibinfo {author}
  {\bibfnamefont {P.}~\bibnamefont {Zoller}}, \ and\ \bibinfo {author}
  {\bibfnamefont {M.}~\bibnamefont {Hafezi}},\ }\href {\doibase
  10.1103/PhysRevX.6.041033} {\bibfield  {journal} {\bibinfo  {journal} {Phys.
  Rev. X}\ }\textbf {\bibinfo {volume} {6}},\ \bibinfo {pages} {041033}
  (\bibinfo {year} {2016})}\BibitemShut {NoStop}%
\bibitem [{Note1()}]{Note1}%
  \BibitemOpen
  \bibinfo {note} {The sign structure of fermionic wave functions makes the
  application of machine learning approaches to fermionic systems more
  involved.}\BibitemShut {Stop}%
\bibitem [{\citenamefont {Torlai}\ \emph {et~al.}(2018)\citenamefont {Torlai},
  \citenamefont {Mazzola}, \citenamefont {Carrasquilla}, \citenamefont
  {Troyer}, \citenamefont {Melko},\ and\ \citenamefont {Carleo}}]{Torlai2018}%
  \BibitemOpen
  \bibfield  {author} {\bibinfo {author} {\bibfnamefont {G.}~\bibnamefont
  {Torlai}}, \bibinfo {author} {\bibfnamefont {G.}~\bibnamefont {Mazzola}},
  \bibinfo {author} {\bibfnamefont {J.}~\bibnamefont {Carrasquilla}}, \bibinfo
  {author} {\bibfnamefont {M.}~\bibnamefont {Troyer}}, \bibinfo {author}
  {\bibfnamefont {R.}~\bibnamefont {Melko}}, \ and\ \bibinfo {author}
  {\bibfnamefont {G.}~\bibnamefont {Carleo}},\ }\href {\doibase
  doi:10.1038/s41567-018-0048-5} {\bibfield  {journal} {\bibinfo  {journal}
  {Nature Physics}\ }\textbf {\bibinfo {volume} {14}},\ \bibinfo {pages}
  {447–450} (\bibinfo {year} {2018})}\BibitemShut {NoStop}%
\bibitem [{\citenamefont {Torlai}\ \emph {et~al.}(2019)\citenamefont {Torlai},
  \citenamefont {Timar}, \citenamefont {van Nieuwenburg}, \citenamefont
  {Levine}, \citenamefont {Omran}, \citenamefont {Keesling}, \citenamefont
  {Bernien}, \citenamefont {Greiner}, \citenamefont {Vuleti{\'c}},
  \citenamefont {Lukin} \emph {et~al.}}]{Torlai2019}%
  \BibitemOpen
  \bibfield  {author} {\bibinfo {author} {\bibfnamefont {G.}~\bibnamefont
  {Torlai}}, \bibinfo {author} {\bibfnamefont {B.}~\bibnamefont {Timar}},
  \bibinfo {author} {\bibfnamefont {E.~P.}\ \bibnamefont {van Nieuwenburg}},
  \bibinfo {author} {\bibfnamefont {H.}~\bibnamefont {Levine}}, \bibinfo
  {author} {\bibfnamefont {A.}~\bibnamefont {Omran}}, \bibinfo {author}
  {\bibfnamefont {A.}~\bibnamefont {Keesling}}, \bibinfo {author}
  {\bibfnamefont {H.}~\bibnamefont {Bernien}}, \bibinfo {author} {\bibfnamefont
  {M.}~\bibnamefont {Greiner}}, \bibinfo {author} {\bibfnamefont
  {V.}~\bibnamefont {Vuleti{\'c}}}, \bibinfo {author} {\bibfnamefont {M.~D.}\
  \bibnamefont {Lukin}},  \emph {et~al.},\ }\href@noop {} {\bibfield  {journal}
  {\bibinfo  {journal} {arXiv preprint arXiv:1904.08441}\ } (\bibinfo {year}
  {2019})}\BibitemShut {NoStop}%
\bibitem [{\citenamefont {Chung}\ \emph {et~al.}(2014)\citenamefont {Chung},
  \citenamefont {Bonnes}, \citenamefont {Chen},\ and\ \citenamefont
  {L\"auchli}}]{Chung2014}%
  \BibitemOpen
  \bibfield  {author} {\bibinfo {author} {\bibfnamefont {C.-M.}\ \bibnamefont
  {Chung}}, \bibinfo {author} {\bibfnamefont {L.}~\bibnamefont {Bonnes}},
  \bibinfo {author} {\bibfnamefont {P.}~\bibnamefont {Chen}}, \ and\ \bibinfo
  {author} {\bibfnamefont {A.~M.}\ \bibnamefont {L\"auchli}},\ }\href {\doibase
  10.1103/PhysRevB.89.195147} {\bibfield  {journal} {\bibinfo  {journal} {Phys.
  Rev. B}\ }\textbf {\bibinfo {volume} {89}},\ \bibinfo {pages} {195147}
  (\bibinfo {year} {2014})}\BibitemShut {NoStop}%
\bibitem [{Note2()}]{Note2}%
  \BibitemOpen
  \bibinfo {note} {The computational complexity of DQMC for simulating the
  total system of $N$ sites is $\sim \beta N^3$ and can be completely decoupled
  from the costly ``exact diagonalization'' inside each HS sample if the
  single-particle Green's functions are saved on disk for every HS
  configuration (or after a number of Monte Carlo steps proportional to the
  autocorrelation time). This requires several hundred GB of hard disk memory
  per parameter set $(\beta , U)$.}\BibitemShut {Stop}%
\bibitem [{\citenamefont {Udagawa}\ and\ \citenamefont
  {Motome}(2010)}]{Udagawa2010}%
  \BibitemOpen
  \bibfield  {author} {\bibinfo {author} {\bibfnamefont {M.}~\bibnamefont
  {Udagawa}}\ and\ \bibinfo {author} {\bibfnamefont {Y.}~\bibnamefont
  {Motome}},\ }\href {\doibase 10.1103/PhysRevLett.104.106409} {\bibfield
  {journal} {\bibinfo  {journal} {Phys. Rev. Lett.}\ }\textbf {\bibinfo
  {volume} {104}},\ \bibinfo {pages} {106409} (\bibinfo {year}
  {2010})}\BibitemShut {NoStop}%
\bibitem [{\citenamefont {Udagawa}\ and\ \citenamefont
  {Motome}(2015)}]{Udagawa2015}%
  \BibitemOpen
  \bibfield  {author} {\bibinfo {author} {\bibfnamefont {M.}~\bibnamefont
  {Udagawa}}\ and\ \bibinfo {author} {\bibfnamefont {Y.}~\bibnamefont
  {Motome}},\ }\href {\doibase 10.1088/1742-5468/2015/01/p01016} {\bibfield
  {journal} {\bibinfo  {journal} {Journal of Statistical Mechanics: Theory and
  Experiment}\ }\textbf {\bibinfo {volume} {2015}},\ \bibinfo {pages} {P01016}
  (\bibinfo {year} {2015})}\BibitemShut {NoStop}%
\bibitem [{\citenamefont {Iglovikov}\ \emph {et~al.}(2015)\citenamefont
  {Iglovikov}, \citenamefont {Khatami},\ and\ \citenamefont
  {Scalettar}}]{Iglovikov2015}%
  \BibitemOpen
  \bibfield  {author} {\bibinfo {author} {\bibfnamefont {V.~I.}\ \bibnamefont
  {Iglovikov}}, \bibinfo {author} {\bibfnamefont {E.}~\bibnamefont {Khatami}},
  \ and\ \bibinfo {author} {\bibfnamefont {R.~T.}\ \bibnamefont {Scalettar}},\
  }\href {\doibase 10.1103/PhysRevB.92.045110} {\bibfield  {journal} {\bibinfo
  {journal} {Phys. Rev. B}\ }\textbf {\bibinfo {volume} {92}},\ \bibinfo
  {pages} {045110} (\bibinfo {year} {2015})}\BibitemShut {NoStop}%
\bibitem [{\citenamefont {Preuss}\ \emph {et~al.}(1995)\citenamefont {Preuss},
  \citenamefont {Hanke},\ and\ \citenamefont {von~der Linden}}]{Preuss1995}%
  \BibitemOpen
  \bibfield  {author} {\bibinfo {author} {\bibfnamefont {R.}~\bibnamefont
  {Preuss}}, \bibinfo {author} {\bibfnamefont {W.}~\bibnamefont {Hanke}}, \
  and\ \bibinfo {author} {\bibfnamefont {W.}~\bibnamefont {von~der Linden}},\
  }\href {\doibase 10.1103/PhysRevLett.75.1344} {\bibfield  {journal} {\bibinfo
   {journal} {Phys. Rev. Lett.}\ }\textbf {\bibinfo {volume} {75}},\ \bibinfo
  {pages} {1344} (\bibinfo {year} {1995})}\BibitemShut {NoStop}%
\bibitem [{\citenamefont {{Huber}}\ \emph {et~al.}(2019)\citenamefont
  {{Huber}}, \citenamefont {{Grusdt}},\ and\ \citenamefont
  {{Punk}}}]{Huber2019}%
  \BibitemOpen
  \bibfield  {author} {\bibinfo {author} {\bibfnamefont {S.}~\bibnamefont
  {{Huber}}}, \bibinfo {author} {\bibfnamefont {F.}~\bibnamefont {{Grusdt}}}, \
  and\ \bibinfo {author} {\bibfnamefont {M.}~\bibnamefont {{Punk}}},\ }\href
  {\doibase 10.1103/PhysRevA.99.023617} {\bibfield  {journal} {\bibinfo
  {journal} {\pra}\ }\textbf {\bibinfo {volume} {99}},\ \bibinfo {eid} {023617}
  (\bibinfo {year} {2019})},\ \Eprint {http://arxiv.org/abs/1808.03653}
  {arXiv:1808.03653 [cond-mat.quant-gas]} \BibitemShut {NoStop}%
\bibitem [{\citenamefont {Blankenbecler}\ \emph {et~al.}(1981)\citenamefont
  {Blankenbecler}, \citenamefont {Scalapino},\ and\ \citenamefont
  {Sugar}}]{Blankenbecler1981}%
  \BibitemOpen
  \bibfield  {author} {\bibinfo {author} {\bibfnamefont {R.}~\bibnamefont
  {Blankenbecler}}, \bibinfo {author} {\bibfnamefont {D.~J.}\ \bibnamefont
  {Scalapino}}, \ and\ \bibinfo {author} {\bibfnamefont {R.~L.}\ \bibnamefont
  {Sugar}},\ }\href {\doibase 10.1103/PhysRevD.24.2278} {\bibfield  {journal}
  {\bibinfo  {journal} {Phys. Rev. D}\ }\textbf {\bibinfo {volume} {24}},\
  \bibinfo {pages} {2278} (\bibinfo {year} {1981})}\BibitemShut {NoStop}%
\bibitem [{\citenamefont {Loh~Jr.}\ and\ \citenamefont
  {Gubernatis}(1992)}]{Loh1992}%
  \BibitemOpen
  \bibfield  {author} {\bibinfo {author} {\bibfnamefont {E.}~\bibnamefont
  {Loh~Jr.}}\ and\ \bibinfo {author} {\bibfnamefont {J.}~\bibnamefont
  {Gubernatis}},\ }in\ \href@noop {} {\emph {\bibinfo {booktitle} {Electronic
  Phase Transitions}}},\ \bibinfo {series} {Modern Problems in Condensed Matter
  Sciences}, Vol.~\bibinfo {volume} {32},\ \bibinfo {editor} {edited by\
  \bibinfo {editor} {\bibfnamefont {W.}~\bibnamefont {Hanke}}\ and\ \bibinfo
  {editor} {\bibfnamefont {Y.}~\bibnamefont {Kopaev}}}\ (\bibinfo  {publisher}
  {North-Holland, Amsterdam},\ \bibinfo {year} {1992})\ Chap.~\bibinfo
  {chapter} {4}, pp.\ \bibinfo {pages} {177--235}\BibitemShut {NoStop}%
\bibitem [{\citenamefont {Assaad}()}]{Assaad2002}%
  \BibitemOpen
  \bibfield  {author} {\bibinfo {author} {\bibfnamefont {F.~F.}\ \bibnamefont
  {Assaad}},\ }in\ \href@noop {} {\emph {\bibinfo {booktitle} {Quantum
  Simulations of Complex Many-Body Systems: From Theory to Algorithms,
  Publication Series of the John von Neumann Institute for Computation
  (NIC)}}}\ (\bibinfo  {publisher} {Edited by J. Grotendorst, D. Marx, and A.
  Muramatsu (NIC, Jülich, 2002)})\BibitemShut {NoStop}%
\bibitem [{\citenamefont {Grover}(2013)}]{Grover2013}%
  \BibitemOpen
  \bibfield  {author} {\bibinfo {author} {\bibfnamefont {T.}~\bibnamefont
  {Grover}},\ }\href {\doibase 10.1103/PhysRevLett.111.130402} {\bibfield
  {journal} {\bibinfo  {journal} {Phys. Rev. Lett.}\ }\textbf {\bibinfo
  {volume} {111}},\ \bibinfo {pages} {130402} (\bibinfo {year}
  {2013})}\BibitemShut {NoStop}%
\bibitem [{Note3()}]{Note3}%
  \BibitemOpen
  \bibinfo {note} {Translational invariance can be used to accumulate
  additional statistics by displacing the subsystem repeatedly. However, for
  the calculations presented in the following for a single plaquette, the
  location of the plaquette was fixed.}\BibitemShut {Stop}%
\bibitem [{\citenamefont {Fano}\ \emph {et~al.}(1992)\citenamefont {Fano},
  \citenamefont {Ortolani},\ and\ \citenamefont {Parola}}]{Fano1992}%
  \BibitemOpen
  \bibfield  {author} {\bibinfo {author} {\bibfnamefont {G.}~\bibnamefont
  {Fano}}, \bibinfo {author} {\bibfnamefont {F.}~\bibnamefont {Ortolani}}, \
  and\ \bibinfo {author} {\bibfnamefont {A.}~\bibnamefont {Parola}},\ }\href
  {\doibase 10.1103/PhysRevB.46.1048} {\bibfield  {journal} {\bibinfo
  {journal} {Phys. Rev. B}\ }\textbf {\bibinfo {volume} {46}},\ \bibinfo
  {pages} {1048} (\bibinfo {year} {1992})}\BibitemShut {NoStop}%
\bibitem [{\citenamefont {Noce}\ and\ \citenamefont {Cuoco}(1996)}]{Noce1996}%
  \BibitemOpen
  \bibfield  {author} {\bibinfo {author} {\bibfnamefont {C.}~\bibnamefont
  {Noce}}\ and\ \bibinfo {author} {\bibfnamefont {M.}~\bibnamefont {Cuoco}},\
  }\href {\doibase 10.1103/PhysRevB.54.13047} {\bibfield  {journal} {\bibinfo
  {journal} {Phys. Rev. B}\ }\textbf {\bibinfo {volume} {54}},\ \bibinfo
  {pages} {13047} (\bibinfo {year} {1996})}\BibitemShut {NoStop}%
\bibitem [{\citenamefont {Schumann}(2002)}]{Schumann2002}%
  \BibitemOpen
  \bibfield  {author} {\bibinfo {author} {\bibfnamefont {R.}~\bibnamefont
  {Schumann}},\ }\href {\doibase
  10.1002/1521-3889(200201)11:1<49::AID-ANDP49>3.0.CO;2-7} {\bibfield
  {journal} {\bibinfo  {journal} {Annalen der Physik}\ }\textbf {\bibinfo
  {volume} {11}},\ \bibinfo {pages} {49} (\bibinfo {year} {2002})}\BibitemShut
  {NoStop}%
\bibitem [{\citenamefont {Kuns}\ \emph {et~al.}(2011)\citenamefont {Kuns},
  \citenamefont {Rey},\ and\ \citenamefont {Gorshkov}}]{Kuns2011}%
  \BibitemOpen
  \bibfield  {author} {\bibinfo {author} {\bibfnamefont {K.~A.}\ \bibnamefont
  {Kuns}}, \bibinfo {author} {\bibfnamefont {A.~M.}\ \bibnamefont {Rey}}, \
  and\ \bibinfo {author} {\bibfnamefont {A.~V.}\ \bibnamefont {Gorshkov}},\
  }\href {\doibase 10.1103/PhysRevA.84.063639} {\bibfield  {journal} {\bibinfo
  {journal} {Phys. Rev. A}\ }\textbf {\bibinfo {volume} {84}},\ \bibinfo
  {pages} {063639} (\bibinfo {year} {2011})}\BibitemShut {NoStop}%
\bibitem [{\citenamefont {Chen}\ and\ \citenamefont
  {Moukouri}(1996)}]{Chen1996}%
  \BibitemOpen
  \bibfield  {author} {\bibinfo {author} {\bibfnamefont {L.}~\bibnamefont
  {Chen}}\ and\ \bibinfo {author} {\bibfnamefont {S.}~\bibnamefont
  {Moukouri}},\ }\href {\doibase 10.1103/PhysRevB.53.1866} {\bibfield
  {journal} {\bibinfo  {journal} {Phys. Rev. B}\ }\textbf {\bibinfo {volume}
  {53}},\ \bibinfo {pages} {1866} (\bibinfo {year} {1996})}\BibitemShut
  {NoStop}%
\bibitem [{\citenamefont {Tinkham}(1964)}]{TinkhamGroupTheory}%
  \BibitemOpen
  \bibfield  {author} {\bibinfo {author} {\bibfnamefont {M.}~\bibnamefont
  {Tinkham}},\ }\href@noop {} {\emph {\bibinfo {title} {Group Theory and
  Quantum Mechanics}}}\ (\bibinfo  {publisher} {McGraw-Hill, New York},\
  \bibinfo {year} {1964})\BibitemShut {NoStop}%
\bibitem [{Note4()}]{Note4}%
  \BibitemOpen
  \bibinfo {note} {Due to the normalization all elements of $\rho _A$ have
  correlations among them so that adding Gaussian noise idependently to all
  elements cannot be entirely correct.}\BibitemShut {Stop}%
\bibitem [{\citenamefont {Scalapino}\ and\ \citenamefont
  {Trugman}(1996)}]{Scalapino1996}%
  \BibitemOpen
  \bibfield  {author} {\bibinfo {author} {\bibfnamefont {D.~J.}\ \bibnamefont
  {Scalapino}}\ and\ \bibinfo {author} {\bibfnamefont {S.~A.}\ \bibnamefont
  {Trugman}},\ }\href {\doibase 10.1080/01418639608240361} {\bibfield
  {journal} {\bibinfo  {journal} {Philosophical Magazine Part B}\ }\textbf
  {\bibinfo {volume} {74}},\ \bibinfo {pages} {607} (\bibinfo {year} {1996})},\
  \Eprint {http://arxiv.org/abs/http://dx.doi.org/10.1080/01418639608240361}
  {http://dx.doi.org/10.1080/01418639608240361} \BibitemShut {NoStop}%
\bibitem [{\citenamefont {Takahashi}(1977)}]{Takahashi1977}%
  \BibitemOpen
  \bibfield  {author} {\bibinfo {author} {\bibfnamefont {M.}~\bibnamefont
  {Takahashi}},\ }\href {http://stacks.iop.org/0022-3719/10/i=8/a=031}
  {\bibfield  {journal} {\bibinfo  {journal} {Journal of Physics C: Solid State
  Physics}\ }\textbf {\bibinfo {volume} {10}},\ \bibinfo {pages} {1289}
  (\bibinfo {year} {1977})}\BibitemShut {NoStop}%
\bibitem [{\citenamefont {MacDonald}\ \emph {et~al.}(1988)\citenamefont
  {MacDonald}, \citenamefont {Girvin},\ and\ \citenamefont
  {Yoshioka}}]{MacDonald1988}%
  \BibitemOpen
  \bibfield  {author} {\bibinfo {author} {\bibfnamefont {A.~H.}\ \bibnamefont
  {MacDonald}}, \bibinfo {author} {\bibfnamefont {S.~M.}\ \bibnamefont
  {Girvin}}, \ and\ \bibinfo {author} {\bibfnamefont {D.}~\bibnamefont
  {Yoshioka}},\ }\href {\doibase 10.1103/PhysRevB.37.9753} {\bibfield
  {journal} {\bibinfo  {journal} {Phys. Rev. B}\ }\textbf {\bibinfo {volume}
  {37}},\ \bibinfo {pages} {9753} (\bibinfo {year} {1988})}\BibitemShut
  {NoStop}%
\bibitem [{\citenamefont {Delannoy}\ \emph {et~al.}(2005)\citenamefont
  {Delannoy}, \citenamefont {Gingras}, \citenamefont {Holdsworth},\ and\
  \citenamefont {Tremblay}}]{Delannoy2005}%
  \BibitemOpen
  \bibfield  {author} {\bibinfo {author} {\bibfnamefont {J.-Y.~P.}\
  \bibnamefont {Delannoy}}, \bibinfo {author} {\bibfnamefont {M.~J.~P.}\
  \bibnamefont {Gingras}}, \bibinfo {author} {\bibfnamefont {P.~C.~W.}\
  \bibnamefont {Holdsworth}}, \ and\ \bibinfo {author} {\bibfnamefont
  {A.-M.~S.}\ \bibnamefont {Tremblay}},\ }\href {\doibase
  10.1103/PhysRevB.72.115114} {\bibfield  {journal} {\bibinfo  {journal} {Phys.
  Rev. B}\ }\textbf {\bibinfo {volume} {72}},\ \bibinfo {pages} {115114}
  (\bibinfo {year} {2005})}\BibitemShut {NoStop}%
\bibitem [{Note5()}]{Note5}%
  \BibitemOpen
  \bibinfo {note} {The colours are based on the sequence of energy levels at
  $|U|/t=4$.}\BibitemShut {Stop}%
\bibitem [{\citenamefont {Keimer}\ \emph {et~al.}(2015)\citenamefont {Keimer},
  \citenamefont {Kivelson}, \citenamefont {Norman}, \citenamefont {Uchida},\
  and\ \citenamefont {Zaanen}}]{Keimer2015}%
  \BibitemOpen
  \bibfield  {author} {\bibinfo {author} {\bibfnamefont {B.}~\bibnamefont
  {Keimer}}, \bibinfo {author} {\bibfnamefont {S.~A.}\ \bibnamefont
  {Kivelson}}, \bibinfo {author} {\bibfnamefont {M.~R.}\ \bibnamefont
  {Norman}}, \bibinfo {author} {\bibfnamefont {S.}~\bibnamefont {Uchida}}, \
  and\ \bibinfo {author} {\bibfnamefont {J.}~\bibnamefont {Zaanen}},\
  }\href@noop {} {\bibfield  {journal} {\bibinfo  {journal} {Nature}\ }\textbf
  {\bibinfo {volume} {518}},\ \bibinfo {pages} {179} (\bibinfo {year}
  {2015})}\BibitemShut {NoStop}%
\bibitem [{\citenamefont {Rey}\ \emph {et~al.}(2009)\citenamefont {Rey},
  \citenamefont {Sensarma}, \citenamefont {F{\"o}lling}, \citenamefont
  {Greiner}, \citenamefont {Demler},\ and\ \citenamefont {Lukin}}]{Rey2009}%
  \BibitemOpen
  \bibfield  {author} {\bibinfo {author} {\bibfnamefont {A.~M.}\ \bibnamefont
  {Rey}}, \bibinfo {author} {\bibfnamefont {R.}~\bibnamefont {Sensarma}},
  \bibinfo {author} {\bibfnamefont {S.}~\bibnamefont {F{\"o}lling}}, \bibinfo
  {author} {\bibfnamefont {M.}~\bibnamefont {Greiner}}, \bibinfo {author}
  {\bibfnamefont {E.}~\bibnamefont {Demler}}, \ and\ \bibinfo {author}
  {\bibfnamefont {M.~D.}\ \bibnamefont {Lukin}},\ }\href
  {http://stacks.iop.org/0295-5075/87/i=6/a=60001} {\bibfield  {journal}
  {\bibinfo  {journal} {EPL (Europhysics Letters)}\ }\textbf {\bibinfo {volume}
  {87}},\ \bibinfo {pages} {60001} (\bibinfo {year} {2009})}\BibitemShut
  {NoStop}%
\bibitem [{\citenamefont {Ying}\ \emph {et~al.}(2014)\citenamefont {Ying},
  \citenamefont {Mondaini}, \citenamefont {Sun}, \citenamefont {Paiva},
  \citenamefont {Fye},\ and\ \citenamefont {Scalettar}}]{Ying2014}%
  \BibitemOpen
  \bibfield  {author} {\bibinfo {author} {\bibfnamefont {T.}~\bibnamefont
  {Ying}}, \bibinfo {author} {\bibfnamefont {R.}~\bibnamefont {Mondaini}},
  \bibinfo {author} {\bibfnamefont {X.~D.}\ \bibnamefont {Sun}}, \bibinfo
  {author} {\bibfnamefont {T.}~\bibnamefont {Paiva}}, \bibinfo {author}
  {\bibfnamefont {R.~M.}\ \bibnamefont {Fye}}, \ and\ \bibinfo {author}
  {\bibfnamefont {R.~T.}\ \bibnamefont {Scalettar}},\ }\href {\doibase
  10.1103/PhysRevB.90.075121} {\bibfield  {journal} {\bibinfo  {journal} {Phys.
  Rev. B}\ }\textbf {\bibinfo {volume} {90}},\ \bibinfo {pages} {075121}
  (\bibinfo {year} {2014})}\BibitemShut {NoStop}%
\bibitem [{\citenamefont {S\'en\'echal}\ \emph {et~al.}(2000)\citenamefont
  {S\'en\'echal}, \citenamefont {Perez},\ and\ \citenamefont
  {Pioro-Ladri\`ere}}]{Senechal2000}%
  \BibitemOpen
  \bibfield  {author} {\bibinfo {author} {\bibfnamefont {D.}~\bibnamefont
  {S\'en\'echal}}, \bibinfo {author} {\bibfnamefont {D.}~\bibnamefont {Perez}},
  \ and\ \bibinfo {author} {\bibfnamefont {M.}~\bibnamefont
  {Pioro-Ladri\`ere}},\ }\href {\doibase 10.1103/PhysRevLett.84.522} {\bibfield
   {journal} {\bibinfo  {journal} {Phys. Rev. Lett.}\ }\textbf {\bibinfo
  {volume} {84}},\ \bibinfo {pages} {522} (\bibinfo {year} {2000})}\BibitemShut
  {NoStop}%
\end{thebibliography}
\end{document}